\documentclass[11pt]{article}
\usepackage[margin=1in]{geometry}
\usepackage{amsmath,amssymb,amsthm,mathtools,microtype,enumitem}
\usepackage[T1]{fontenc}
\usepackage{lmodern}
\usepackage{booktabs,tabularx}
\usepackage{needspace}
\usepackage{xcolor}
\usepackage[colorlinks=true,linkcolor=blue!60!black,citecolor=blue!60!black,urlcolor=blue!60!black]{hyperref}
\usepackage[capitalize,noabbrev]{cleveref}
\usepackage{tikz}
\usetikzlibrary{arrows.meta}
\hypersetup{pdftitle={Near-linear-Time Sampling of the Ferromagnetic Ising Model at the Tree Uniqueness Threshold},
pdfauthor={Tianshun Miao},
pdfsubject={Ising, XY, and Potts sampling through flow and polymer representations}}
\newtheorem{theorem}{Theorem}[section]
\newtheorem{lemma}[theorem]{Lemma}
\newtheorem{proposition}[theorem]{Proposition}
\newtheorem{corollary}[theorem]{Corollary}
\theoremstyle{definition}\newtheorem{definition}[theorem]{Definition}
\newtheorem{example}[theorem]{Example}
\theoremstyle{remark}
\theoremstyle{definition}
\theoremstyle{remark}
\crefname{introdefinition}{Definition}{Definitions}
\Crefname{introdefinition}{Definition}{Definitions}
\crefname{introremark}{Remark}{Remarks}
\Crefname{introremark}{Remark}{Remarks}
\theoremstyle{plain}
\newcommand{\wtO}{\widetilde O}
\newcommand{\tmix}{t_{\mathrm{mix}}}

\newcommand{\E}{\mathbb E}
\newcommand{\Prb}{\mathbb P}
\newcommand{\Var}{\operatorname{Var}}
\newcommand{\Cov}{\operatorname{Cov}}
\newcommand{\Ent}{\operatorname{Ent}}
\newcommand{\Diag}{\operatorname{Diag}}

\newcommand{\cE}{\mathcal E}

\newcommand{\one}{\mathbf 1}

\newcommand{\TV}{d_{\mathrm{TV}}}
\DeclareMathOperator{\gap}{gap}
\newcommand{\od}{\mathrm{od}}
\setlist{itemsep=3pt,topsep=5pt}
\title{Near-linear-Time Sampling of the Ferromagnetic Ising Model at the Tree Uniqueness Threshold}
\author{Tianshun Miao\\[0.6em]
\normalsize State Key Laboratory for Novel Software Technology,\\
\normalsize New Cornerstone Science Laboratory,\\
\normalsize Nanjing University, China\\[0.3em]
\normalsize\texttt{miaotianshun@smail.nju.edu.cn}}
\date{}
\begin{document}
\hypersetup{pageanchor=false}
\pagenumbering{gobble}
\maketitle

\begin{abstract}
We give a near-linear-time sampler for the zero-field ferromagnetic
Ising model throughout the tree-uniqueness regime, including the
threshold itself. The method also gives sampling algorithms for the
XY and integer Potts models. On an $n$-vertex graph with $m$ edges and
maximum degree at most $\Delta$, we obtain the following expected-work
bounds, with total variation error at most $\varepsilon$.
\begin{enumerate}[label=(\roman*),leftmargin=*,itemsep=4pt,topsep=5pt]
\item \textbf{Near-linear Ising sampling.}
For $\Delta\ge3$, the work is $O(n+m\log^3(n/\varepsilon))$
throughout the tree-uniqueness regime, including the threshold
$\beta=\Delta/(\Delta-2)$.
\item \textbf{Almost-linear XY sampling.}
For $\Delta\ge8$, we sample the XY model in
$\widetilde O(m+n^{1+o(1)})$ expected work at inverse-polynomial accuracy
up to and including the robust phase-transition threshold on the
$\Delta$-regular tree.
\item \textbf{Near-linear Potts sampling.}
For integer $q\ge3$, we sample the $q$-state Potts model in
$\widetilde O(n+q^{O(1)}m)$ expected work whenever
$\beta-1\le 0.1(\beta_{\od}(q,\Delta)-1)$, where $\beta_{\od}$ is the
ordered/disordered coexistence parameter.
\end{enumerate}
We express the even-subgraph representation of critical Ising as a
polymer model that admits near-linear-time sampling, thereby bypassing
the critical slowdown of Glauber and Swendsen--Wang dynamics.
Building on the low-degree approximation approach of Miao and Yin,
we establish general spectral conditions for fast polymer sampling.
We verify these conditions through explorations that also implement
the sampler efficiently, and apply the same framework to flow
representations of the XY and Potts models.
We also establish polynomial-time counting and sampling for the
ferromagnetic XY model at every fixed temperature.
\end{abstract}
\clearpage
{\small\setlength{\parskip}{0pt}
\tableofcontents
}
\clearpage
\pagenumbering{arabic}
\hypersetup{pageanchor=true}

\section{Introduction}\label{sec:intro}

The ferromagnetic Ising model is a basic model of magnetism and phase
transitions in statistical physics. Each vertex carries a binary spin,
with neighboring spins favoring agreement. Sampling its Gibbs
distribution gives access to typical configurations and physical
observables such as magnetization and spin correlations \cite{RW99}.
The model has also inspired research in discrete probability,
machine learning, and theoretical computer science.
Polynomial-time approximate counting and sampling of this model have been known
since the work of Jerrum and Sinclair \cite{JSIsing93} and Randall and
Wilson \cite{RW99}. A more refined question is how small the exponent
in the polynomial running time can be.

We consider simple graphs $G=(V,E)$ with $n$ vertices, $m$ edges, and
maximum degree at most $\Delta$.
For $\sigma\in\{-1,1\}^V$, let $m_{=}(\sigma)$ count edges whose
endpoints have equal spins. The zero-field ferromagnetic Ising law is defined as
\[
 \mu_{\rm Ising}(\sigma)=Z_{\rm Ising}^{-1}\beta^{m_{=}(\sigma)},
 \qquad \beta\ge1.
\]
Here $Z_{\rm Ising}$ is the normalizing constant, called the partition
function. In this parametrization, the uniqueness threshold on the infinite
$\Delta$-regular tree is
$\beta_{\rm Ising}(\Delta)=\Delta/(\Delta-2)$ for $\Delta\ge3$.
Strictly below this threshold, for every fixed $\xi\in(0,1)$,
Glauber dynamics mixes to total-variation error $\varepsilon$ in
$O_\xi(n\log(n/\varepsilon))$ updates when
$1<\beta\le(\Delta-\xi)/(\Delta-2+\xi)<\beta_{\rm Ising}(\Delta)$
\cite{CFYZ21Ising}.

At this threshold, however, standard Markov chains can require
superlinear work.
Single-site Glauber dynamics can require $\Omega(n^{3/2})$ updates
on bounded-degree graphs at the tree threshold \cite{CCYZ25}.
Swendsen--Wang dynamics also exhibits critical slowdown: its mixing
time is $\Theta(n^{1/4})$ rounds in the critical mean-field model
\cite{LNNP14SW}. For sparse graphs, we give a simple
$\Omega(n^{1/8})$-round lower bound on some cubic graphs at their
tree threshold (\cref{sw:prop:cubic}). Ordinary full sweeps cost
$\Theta(n)$ each on these graphs, giving an $\Omega(n^{9/8})$
worst-start mixing cost. We call the zero-field model at $\beta=\beta_{\rm Ising}(\Delta)$
\emph{critical Ising} and study its worst-case sampling complexity
on finite graphs. These observations leave the following question:
\begin{quote}
\emph{Can critical Ising be sampled in time near-linear in the input size
on every graph of maximum degree at most $\Delta$?}
\end{quote}

We answer this question affirmatively, showing that the critical
slowdown of these dynamics can be bypassed using a different sampling scheme.
The same approach also yields sampling algorithms for the XY and
integer Potts models. In each case, we sample a suitable polymer
representation and then recover the original spins;
\cref{sec:proof-strategy} explains the conditions that make these
representations easy to sample.

\subsection{Sampling guarantees}\label{sec:sampling-guarantees}

An $\varepsilon$-approximate sample is within total-variation distance
$\varepsilon$ of the target law. The notation $\widetilde O$ suppresses
polylogarithmic factors in $n/\varepsilon$. All work bounds include
preprocessing and the output of all $n$ spins.

\paragraph{Ising at the tree-uniqueness threshold.}

Our sampler achieves near-linear work on every graph of maximum
degree at most $\Delta$ throughout the tree-uniqueness regime,
including its critical endpoint.

\begin{theorem}[Ising sampling at the tree-uniqueness threshold]\label{spin:thm:ising}
Let $G$ have maximum degree at most $\Delta\ge3$.
If $1<\beta\le\Delta/(\Delta-2)$, then, for every $0<\varepsilon<1/2$,
the zero-field Ising law has an $\varepsilon$-approximate sampler with
expected work $O(n+m\log^3(n/\varepsilon))$.
\end{theorem}

We express the even-subgraph representation of critical Ising as a
polymer model and prove that it satisfies the conditions for efficient
sampling described in \cref{sec:proof-strategy}.
Sampling this representation and then recovering the spins allows us
to bypass the difficulties of sampling spin configurations directly.

\paragraph{XY at the robust tree threshold.}
The XY model extends binary spins to continuous spins taking values
on the unit circle. Ising and XY are the $N=1$ and $N=2$ cases of
the classical $O(N)$ spin model of statistical physics, respectively
\cite{LiuZhangTanDeng23}.
On two-dimensional lattices, the XY model is a standard setting
for the study of vortex-driven phase transitions
\cite{KT73Ordering}.
Here we study its zero-field Gibbs law on general graphs:
\[
 \mu_{\rm XY}(d\theta)=Z_{\rm XY}^{-1}
 \beta^{\sum_{uv\in E}\cos(\theta_u-\theta_v)}
 \prod_{v\in V}\frac{d\theta_v}{2\pi},
 \qquad \theta\in[0,2\pi)^V.
\]
Its robust phase-transition threshold on the infinite
$\Delta$-regular tree is the unique $\beta_{\rm XY}(\Delta)>1$ satisfying
\begin{equation}\label{spin:eq:xy-range}
 (\Delta-1)\frac{I_1(\log\beta_{\rm XY}(\Delta))}
                     {I_0(\log\beta_{\rm XY}(\Delta))}=1,
\end{equation}
where $I_k(x)=\frac1{2\pi}\int_0^{2\pi}e^{x\cos\theta}\cos(k\theta)\,d\theta$
is the modified Bessel function of the first kind.
Above the robust threshold, an aligned distant boundary can bias the
root even when its connections to the interior are made arbitrarily
weak. Below it, sufficiently weak boundary connections make that
influence vanish \cite{PemantleSteif99}.
This is a physical threshold on the infinite tree; our algorithmic
guarantee below concerns every finite graph of maximum degree at most
$\Delta$.

Bauerschmidt and Bodineau \cite{BB19SpinLSI} establish a
log--Sobolev inequality in the high-temperature range
$\Delta\log\beta<1$, yielding exponential convergence in relative
entropy for continuous-spin Langevin dynamics.
Using our polymer sampling framework, we extend the range of efficient
XY sampling up to and including the robust tree threshold, with an
almost-linear work bound.

\begin{theorem}[XY sampling at the robust tree threshold]\label{spin:thm:xy}
Let $G$ have maximum degree at most $\Delta\ge8$.
If $1<\beta\le\beta_{\rm XY}(\Delta)$, then, for every
$0<\varepsilon<1/2$, the XY law has an $\varepsilon$-approximate sampler
with expected work
\[
 \widetilde O\!\left(m+n\exp\!\left(O\!\left(\sqrt{\log(n/\varepsilon)}\right)\right)\right).
\]
The work bound uses the real-arithmetic model of
\cref{sec:computational-model}.
\end{theorem}

For $\varepsilon=n^{-O(1)}$, this is
$\widetilde O(m+n^{1+o(1)})$ work.
\Cref{spin:app:permanent} gives polynomial-time counting and sampling
for XY at every fixed temperature, including beyond the robust tree
threshold.
\Cref{spin:thm:xy} supplies the stronger almost-linear guarantee
throughout the stated range.

\paragraph{Potts below the order--disorder scale.}
The ferromagnetic $q$-state Potts model generalizes Ising to more than two
equivalent spin states, with neighboring spins favoring agreement.
It is a basic model for the competition between disorder and
symmetry-related ordered phases \cite{Wu82Potts}.
For an integer $q\ge3$, its zero-field law is
$\mu_{q,\beta}(\sigma)=Z_{q,\beta}^{-1}\beta^{m_{=}(\sigma)}$ on
$\sigma\in[q]^V$. Here $m_{=}(\sigma)$ counts edges whose endpoints
have equal spins, as in the Ising model. Define
\[
 \beta_{\od}(q,\Delta)=
 \frac{q-2}{(q-1)^{1-2/\Delta}-1},\qquad \Delta\ge3.
\]
The parameter $\beta_{\od}$ is called the order--disorder threshold
because the ordered and disordered phases on the infinite
$\Delta$-regular tree have equal Bethe free energy at this value.
For $\beta<\beta_{\od}$, the disordered phase is dominant in the
Bethe description \cite[Theorem~4]{GSVY16Potts}, suggesting that
efficient counting and sampling should be possible throughout this
high-temperature regime.
For $\beta>\beta_{\od}$, Galanis et al.\
\cite[Theorem~2]{GSVY16Potts} show that approximate counting is
$\#\mathrm{BIS}$-hard, even on
bipartite graphs of maximum degree $\Delta$.

On the high-temperature side, Carlson et al.\ \cite{CDFKPY24Potts}
give polynomial-time counting and sampling algorithms on all graphs
of maximum degree $\Delta$ when
$\log\beta\le(1-\xi)\log\beta_{\od}(q,\Delta)$, for fixed $\xi>0$
and $\Delta,q$ sufficiently large in terms of $\xi$.
For fixed such $\Delta$ and growing $q$, this upper endpoint is
$(1+o(1))q^{2(1-\xi)/\Delta}$, whereas
$\beta_{\od}(q,\Delta)=(1+o(1))q^{2/\Delta}$.
We obtain near-linear sampling up to a constant fraction of
$\beta_{\od}-1$, for every $q,\Delta\ge3$.

\begin{theorem}[Potts sampling below coexistence]\label{spin:thm:potts}
Let $q\ge3$ be an integer and let $G$ have maximum degree at most
$\Delta\ge3$. If
\[
 0\le\beta-1\le0.1\bigl(\beta_{\od}(q,\Delta)-1\bigr),
\]
then, for every $0<\varepsilon<1/2$, the zero-field Potts law has an
$\varepsilon$-approximate sampler with expected work
$\widetilde O(n+q^{O(1)}m)$. The exponent of $q$ is absolute and independent
of $\Delta$. The same bound holds for the random-cluster law with edge
activity $\beta-1$ and cluster weight $q$.
\end{theorem}

Write $\beta_* = 1+0.1(\beta_{\od}-1)$ for our upper endpoint.
For fixed $\Delta$ and $q\to\infty$, it satisfies
$\beta_*=(0.1+o(1))q^{2/\Delta}$.
Thus the guarantee reaches the order--disorder scale within a
constant factor in the edge activity.
In the inverse-temperature variable $\log\beta$ used by
\cite{CDFKPY24Potts}, this multiplicative approximation becomes
an additive one:
\[
 0\le\log\beta_{\od}-\log\beta_*\le\log 10.
\]

\subsection{Proof strategy}\label{sec:proof-strategy}

The Ising result combines a spectral estimate that remains strict at
criticality with a cycle sampler whose proposals can be generated
efficiently. We first state the general polymer theorem that turns
these estimates into a mixing bound, then explain how explorations
verify its hypotheses and implement its updates. The XY and Potts
applications use the same theorem with different representations.

\begin{definition}[Polymer model]\label{pol:def:model}
A polymer model consists of a type space $\mathcal P$, positive
activities $w_\gamma$, and a symmetric compatibility relation $\sim$,
with $\gamma\not\sim\gamma$ for every type. Its configurations are
finite pairwise compatible families, with law
\[
 \mu(X)=\frac1Z\prod_{\gamma\in X}w_\gamma,
 \qquad
 Z=\sum_{X\text{ compatible}}\prod_{\gamma\in X}w_\gamma.
\]
\end{definition}
This is the hard-core model on the incompatibility graph; the polymer
types carry the combinatorial structure of the application
\cite[Section~1.1]{LiuPolymerNotes23}.

\paragraph{Fast sampling under spectral conditions.}
The three applications follow from a general sampling theorem for
polymer models.
Write $A_{\gamma\gamma'}=\one_{\{\gamma\not\sim\gamma'\}}$,
$W=\Diag(w)$, and
\[
 D(\gamma)=\sum_{\gamma'\not\sim\gamma}w_{\gamma'}.
\]
Thus $D(\gamma)$ measures the total activity conflicting with $\gamma$.
Spectral bounds on $AW$ refer to the symmetric matrix
$W^{1/2}AW^{1/2}$.

\begin{theorem}[Polymer sampling]\label{pol:lem:sampling}
Let $N\ge2$ be an integer such that
\begin{equation}\label{pol:eq:global-bounds}
 \sum_\gamma w_\gamma\le N,
 \qquad |X|\le N\quad\text{for every compatible family }X.
\end{equation}
Suppose that, for a fixed $\eta>0$,
\begin{equation}\label{pol:eq:activity-tail}
 \sum_\gamma w_\gamma e^{\eta D(\gamma)}\le N.
\end{equation}
Consider the insertion--deletion chain with the following transition
probabilities from a configuration $X$:
\begin{enumerate}[label=(\roman*)]
\item For each polymer $\gamma$ compatible with $X$, move to
$X\cup\{\gamma\}$ with probability $w_\gamma/(4N)$.
\item For each $\gamma\in X$, move to $X\setminus\{\gamma\}$ with
probability $1/(4N)$.
\item Stay at $X$ with the remaining probability.
\end{enumerate}
For fixed $0<\delta<1$ and $0<\varepsilon<1/2$, starting from the empty
family gives a sample within total variation distance $\varepsilon$
of $\mu$ in the following numbers of steps:
\begin{center}
\begin{tabular}{@{}ll@{}}
\toprule
Spectral hypothesis & Chain steps\\
\midrule
$\lambda_{\min}(AW)\ge-1+\delta$
 & $O\!\left(N\log^2\frac N\varepsilon\right)$\\[4pt]
$\lambda_{\max}(AW)\le e-\delta$
 & $N\exp\!\left(O\!\left(\sqrt{\log\frac N\varepsilon}\right)\right)$\\
\bottomrule
\end{tabular}
\end{center}
The constants depend only on $\eta$ and $\delta$.
\end{theorem}

\phantomsection\label{pol:rem:continuous}
The theorem also applies to continuously indexed types with activity
measure $\omega$: activity sums become integrals, insertion proposals
have measure $d\omega/(4N)$, and $AW$ is the conflict operator on
$L^2(\omega)$ defined in \cref{pol:sec:definitions}.
The size parameter $N$ bounds the total activity and configuration size,
allowing the theorem to apply to exponentially large or continuous
type spaces.
Its conclusion counts chain steps; the application-specific proposal
and compatibility procedures determine the work per step.

The proof compares insertion--deletion dynamics started at the empty
family with a finite random pool, building on the random discretization
of repulsive Gibbs point processes by
Friedrich et al.\ \cite{FGKKP25Random}.
The activity tail permits truncation of rare types. Spectral
preservation under sampling \cite{SW25Sampling} controls the conflict
matrix of the Poisson-discretized model, which also supplies
occupied-marginal lower bounds.

The finite-model analysis uses the log--Sobolev criterion obtained by
Miao and Yin via low-degree approximation
\cite[Theorem~1.4]{MY26Matchings}, which allows polynomially small
occupied marginals at only a logarithmic cost.
The lower-spectrum hypothesis gives a bounded covariance integral
through the field-dynamics trickle-down theorem of
Chen et al.\ \cite{CCCYZ25}; the upper-spectrum hypothesis gives a
weaker integral bound through a potential-function argument.
A comparison of finite trajectories transfers these bounds to the
original chain. The implementation proposes polymers directly in
their original space, without constructing the pool or paying for
its number of coordinates.
For Ising, first-hit exploration alone bounds the weighted conflict
spectrum by $[-1,1]$, whereas the near-linear branch requires a strict
margin above $-1$. The completion-event construction described below
improves the lower bound to $-5/6$, retaining a margin of $1/6$ at
criticality.
For Potts, the upper bound $\lambda_{\max}(AW)\le1/2$ also gives
$\lambda_{\min}(AW)\ge-1/2$, so the same branch applies.
For XY, the exploration gives
\[
 \lambda_{\max}(AW)\le
 \frac{2\Delta^2(\Delta-2)}{\Delta(\Delta-2)^2-2}
 \qquad\text{when }\beta\le\beta_{\rm XY}(\Delta)
\]
by \cref{spin:lem:xy-ratios,spin:prop:xy-hit}.
This upper bound decreases with $\Delta$ and is strictly below $e$
for $\Delta\ge8$. The upper-spectrum branch therefore gives
almost-linear work in this degree range, explaining the restriction
in \cref{spin:thm:xy}.

\paragraph{Graph explorations for spectral bounds and sampling.}
Random walks and related explorations of the underlying graph serve
two purposes. First, by tracking when an exploration first reaches
a target polymer and whether it subsequently completes that polymer,
we bound $\lambda_{\max}(AW)$ and the total polymer activity
(\cref{pol:lem:first-hit}). Further estimates for completion events
give the lower spectral bound and activity control needed by the
Ising and Potts applications (\cref{spin:lem:event-bounds}).

Second, with suitable acceptance corrections, these explorations
provide the proposal oracle for the insertion--deletion sampler in
\cref{pol:lem:sampling}. Each trial produces a candidate with measure
$d\omega/(4N)$, or returns no candidate. Thus we can generate proposals
with exactly the required activity weights without enumerating the
polymer space. The same exploration approach supplies both the
estimates used to prove rapid mixing and the proposals used to
implement the sampler.

\paragraph{Exact representations and efficient proposals.}
We construct polymer representations and
exact proposals that work throughout the stated temperature ranges.
For Ising, a degree reduction turns the even-subgraph representation
into a model of cycle polymers. A nonbacktracking exploration proposes
cycles, and a root correction gives each cycle its prescribed activity. Its harmonic
length bound controls unsuccessful trials as well as successful ones
(\cref{spin:lem:ising-oracle}).
The XY representation retains the full Bessel weights,
including all integer-flow multiplicities. Continuous interval marks
turn the resulting interactions into pairwise hard-core constraints
(\cref{spin:prop:xy-projection}). A last-exit correction removes the
multiplicity of Euler-word proposals and yields the prescribed polymer
activity before drawing the marks (\cref{spin:lem:xy-oracle}).
Potts proposals include a test implemented by a small regional random-cluster
sample. The work bounds include failed trials, compatibility tests,
and recovery of all spins.

\subsection{Related work}\label{sec:related-work}

\paragraph{Ising, XY, and \texorpdfstring{$q$}{q}-Potts sampling.}
For zero-field Ising, Feng, Guo, and Zhang
\cite[Theorem~1]{FGZ26ZeroField} obtain sampling time
$O(m\log^2 n+n)+O_\beta(n^2\log^2 n\log^2(n/\varepsilon))$ for every
fixed $\beta>1$. Within the strict tree-uniqueness regime, Chen,
Feng, Yin, and Zhang \cite[Theorem~1.1]{CFYZ21Ising} prove
$O_\xi(n\log(n/\varepsilon))$ Glauber mixing when
$1<\beta\le(\Delta-\xi)/(\Delta-2+\xi)$, for fixed $\xi\in(0,1)$.
At the endpoint, Chen, Chen, Yin, and Zhang \cite{CCYZ25} establish
polynomial mixing and the $\Omega(n^{3/2})$ lower bound discussed
above. Our theorem gives near-linear sampling throughout the
tree-uniqueness regime with a bound uniform up to its endpoint.
Near-linear sampling with consistently biased external fields is
also known \cite{CZ23IsingField}; that result requires the field
activities to be bounded away from one and hence excludes zero field.

For continuous spins, Bauerschmidt and Bodineau
\cite[Theorem~1]{BB19SpinLSI} prove a log--Sobolev inequality under a
spectral condition on the interaction matrix. In our XY
parametrization, their condition is
$(\log\beta)(\lambda_{\max}(A_G)-\lambda_{\min}(A_G))<2$, where
$A_G$ is the graph adjacency matrix. The bound
$\lambda_{\max}(A_G)-\lambda_{\min}(A_G)\le2\Delta$ gives the
sufficient condition $\Delta\log\beta<1$ used in
\cref{sec:sampling-guarantees}. This controls entropy decay for the
associated continuous-spin diffusion. Graphical representations also
support worm algorithms for $O(N)$ models
\cite{LiuZhangTanDeng23}, studied there through numerical simulations.
Our result gives a worst-case almost-linear work bound for an
implemented XY sampler up to the robust tree threshold.

For Potts, the high-temperature result of Carlson et al.\
\cite{CDFKPY24Potts} applies to all graphs of maximum degree $\Delta$.
Their counting algorithm
runs in $(n/\varepsilon)^{O_\xi(\log\Delta)}$ time in the range
described in \cref{sec:sampling-guarantees}. Our sampler has
near-linear dependence on the graph size for every $q,\Delta\ge3$
and reaches a constant-factor approximation to $\beta_{\od}$,
or an additive-constant approximation to $\log\beta_{\od}$.
Flow representations have also been used for low-temperature Potts
sampling by Huijben, Patel, and Regts \cite{HPR23Flows}.
We sample connected flow components in the high-temperature regime
using the same $\mathbb Z_q$-flow and random-cluster correspondence.

\paragraph{Polymer sampling and conditions.}
For subset polymers on an $n$-vertex host graph, write $d$ for its
maximum degree, $s$ for the number of permitted labels per vertex,
and $|\gamma|$ for the support size.
Chen et al.\ \cite[Definitions~1 and~4, Theorem~5]{CGGGPSV21Polymer}
analyze insertion--deletion dynamics under the mixing condition
\[
 \sum_{\gamma'\not\sim\gamma}|\gamma'|w_{\gamma'}
 \le\theta|\gamma|\qquad\text{for every }\gamma,
 \quad \theta<1.
\]
For efficient proposals, they impose the stronger sampling condition
\[
 w_\gamma\le e^{-\tau|\gamma|}
 \qquad\text{with}\qquad \tau\ge5+3\log(sd).
\]
When polymer eligibility and weights can be computed in time
polynomial in $|\gamma|$, this gives approximate sampling in
$O(n\log(n/\varepsilon)\log(1/\varepsilon))$ time.

Friedrich et al.\ \cite[Condition~1, Theorem~2]{FGKP23Cliques}
use clique dynamics under the condition that some positive function
$f$ satisfies
\[
 \sum_{\substack{\gamma'\not\sim\gamma\\\gamma'\ne\gamma}}
 f(\gamma')\frac{w_{\gamma'}}{1+w_{\gamma'}}
 \le f(\gamma)\qquad\text{for every }\gamma.
\]
For a cover by $k$ cliques of mutually incompatible polymers, their
sampling theorem additionally assumes computational feasibility,
$e^{-\operatorname{poly}(k)}\le f(\gamma)\le
e^{\operatorname{poly}(k)}$,
\[
 \max_{i\in[k]}\left(1+\sum_{\gamma\in\Lambda_i}w_\gamma\right)
 \le\operatorname{poly}(k),
\]
and a polynomial-time sampler for the polymer distribution restricted
to each clique $\Lambda_i$. These assumptions give sampling time
polynomial in $k/\varepsilon$.

Blanca, Cannon, and Perkins
\cite[Theorem~1.4]{BCP22Perfect} obtain perfect sampling in expected
$O(n\log n)$ time for subset polymer models with
\[
 w_\gamma\le\lambda^{|\gamma|},\qquad
 \lambda<\frac{(d-2)^{d-2}}{s(d-1)^{d-1}},\qquad
 \sum_{\gamma\not\sim v}|\gamma|w_\gamma\le\theta<1
 \quad\text{for every }v.
\]
Here $\gamma\not\sim v$ means that the support of $\gamma$ meets the
closed neighborhood of $v$. The theorem fixes $d\ge3$, $s\ge1$,
$\lambda$, and $\theta$, assumes polynomial-time weight evaluation
in $|\gamma|$, and requires a fixed $\kappa>0$ such that every vertex
belongs to some polymer of weight at least $\kappa$.
We use the stated full versions of \cite{FGKP23Cliques,BCP22Perfect},
as specified in the bibliography.

To compare these conditions with \cref{pol:lem:sampling}, let
$\ell_\gamma=|\gamma|$. The mixing condition of
\cite{CGGGPSV21Polymer} is $AW\ell\le\theta\ell$, which implies
$\lambda_{\max}(AW)\le\theta<1$. The clique condition controls
the off-diagonal incompatibility matrix with weights
$w_\gamma/(1+w_\gamma)$.
Our theorem uses an activity tail in the total conflicting weight
$D(\gamma)$, together with either
$\lambda_{\min}(AW)\ge-1+\delta$ for near-linear mixing or
$\lambda_{\max}(AW)\le e-\delta$ for almost-linear mixing.
It therefore permits upper eigenvalues exceeding one and does not
require pointwise exponential decay in support size.
At critical Ising, the lower-spectrum bound retains the strict margin
needed for the near-linear conclusion. In each application, efficient
proposals must still be constructed separately to turn the mixing
bound into a running-time guarantee.

\paragraph{Organization.}
\Cref{sec:preliminaries} collects the Markov-chain preliminaries,
model representations, and computational conventions.
\Cref{pol:sec:sampling} proves the shared exploration estimates and
\cref{pol:lem:sampling}. The Ising, XY, and Potts constructions, including
their implementation and work bounds, are proved in
\cref{spin:sec:ising,spin:sec:xy,spin:sec:potts}.
\Cref{sec:discussion} discusses extensions and open problems.
\Cref{spin:app:permanent} gives the general-temperature XY reduction.
\Cref{sw:app:lower-bound} proves the Swendsen--Wang lower bound on
critical cubic graphs.

\section{Preliminaries}\label{sec:preliminaries}
All logarithms are natural unless a base is indicated.

\subsection{Markov chains and mixing}\label{sec:markov-preliminaries}

For probability laws $\rho,\pi$ on the same measurable space $\Omega$,
their \emph{total-variation distance} is
\[
 \TV(\rho,\pi)=\sup_{A\subseteq\Omega\text{ measurable}}
 |\rho(A)-\pi(A)|.
\]
On a finite or countable space, this equals
$\frac12\sum_{x\in\Omega}|\rho(x)-\pi(x)|$.
Every Markov kernel $K$ contracts total variation:
$\TV(\rho K,\pi K)\le\TV(\rho,\pi)$.

Let $P$ be an irreducible and aperiodic Markov kernel on a finite
state space $\Omega$, with stationary law $\pi$, so $\pi P=\pi$.
We write $\rho P^t$ for the law after $t$ transitions from initial
law $\rho$. The kernel is \emph{reversible} if
$\pi(x)P(x,y)=\pi(y)P(y,x)$ for all $x,y\in\Omega$, and is
\emph{lazy} if $P(x,x)\ge1/2$ for all $x$.
For $0<\varepsilon<1$, its worst-start \emph{mixing time} is
\[
 \tmix(P,\varepsilon)
 =\min\left\{t\in\mathbb Z_{\ge0}:
 \max_{x\in\Omega}\TV(P^t(x,\cdot),\pi)\le\varepsilon\right\}.
\]
For general state spaces, the same definition uses a supremum over
initial states, and the mixing time may be infinite.
Time counts transitions of the specified kernel; the work needed
to implement each transition is accounted for separately.
The guarantee in \cref{pol:lem:sampling} concerns the particular
initial state $X=\varnothing$, rather than the worst initial state.

\subsection{Dirichlet forms, entropy, and spectral gap}
\label{sec:functional-inequalities}

Let $P$ be an irreducible $\pi$-reversible Markov kernel on a finite
state space $\Omega$ with at least two states. For a real function
$f$ and a nonnegative function $h$, define the \emph{Dirichlet form}
and \emph{entropy} by
\begin{align*}
 \cE_P(f,f)
 &=\frac12\sum_{x,y\in\Omega}\pi(x)P(x,y)
       (f(x)-f(y))^2,\\
 \Ent_\pi(h)
 &=\E_\pi[h\log h]-\E_\pi[h]\log\E_\pi[h],
\end{align*}
with $0\log0=0$.
A \emph{log--Sobolev inequality} (LSI) with constant $\alpha>0$ is
\[
 \Ent_\pi(f^2)\le\alpha^{-1}\cE_P(f,f)
 \qquad\text{for all }f:\Omega\to\mathbb R.
\]
We obtain this inequality from \cite[Theorem~1.4]{MY26Matchings},
as specialized in \cref{pol:lem:external-lsi}; see \cite{DSC96}
for background.
For a probability law $\rho$ with density $h=\rho/\pi$,
Pinsker's inequality gives
\begin{equation}\label{mc:eq:pinsker}
 \TV(\rho,\pi)^2\le\frac12\Ent_\pi(h).
\end{equation}

Writing $\Var_\pi(f)=\E_\pi[f^2]-(\E_\pi f)^2$, the
\emph{spectral gap} is
\[
 \gap(P)=\inf_{\Var_\pi(f)>0}
 \frac{\cE_P(f,f)}{\Var_\pi(f)}=1-\lambda_2(P),
\]
where the eigenvalues are ordered algebraically.
Equivalently, $\Var_\pi(f)\le\gap(P)^{-1}\cE_P(f,f)$ is the
Poincar\'e inequality.
For an aperiodic $P$, the spectral lower bound
\cite[Theorem~12.5]{LP17Markov} gives
\begin{equation}\label{mc:eq:gap-lower}
 \tmix(P,\varepsilon)
 \ge\bigl(\gap(P)^{-1}-1\bigr)\log\frac1{2\varepsilon},
 \qquad 0<\varepsilon<1/2.
\end{equation}
This is the implication used in \cref{sw:app:lower-bound}.

\subsection{Spin models and flow representations}\label{spin:sec:flows}

For edge activities $\beta_e\ge1$, the zero-field Ising law is
\[
 \mu_{\rm Ising}(\sigma)=\frac1{Z_{\rm Ising}}
 \prod_{e=uv\in E}\beta_e^{\one_{\{\sigma_u=\sigma_v\}}},
 \qquad \sigma\in\{-1,1\}^V.
\]
Writing $t_e=(\beta_e-1)/(\beta_e+1)$, its even-subgraph representation is

\begin{equation}\label{spin:eq:ising-flow}
 \pi_2(F)=\frac1{Z_2}
 \one_{\{\deg_F(v)\text{ is even for every }v\}}
 \prod_{e\in F}t_e,
 \qquad F\subseteq E,
\end{equation}
with $Z_{\rm Ising}=2^n\prod_e[(\beta_e+1)/2]\,Z_2$.
Equivalently, $\one_F$ is a $\mathbb Z_2$-flow. We also use
\eqref{spin:eq:ising-flow} directly with edge weights $0<t_e\le1$.
See \cite{GrimmettJanson09} for the even-subgraph representation.

The XY law at parameter $\beta>1$ is
\[
 \mu_\beta(d\theta)=\frac1{Z_{\rm XY}}
 \beta^{\sum_{uv\in E}\cos(\theta_u-\theta_v)}
 \prod_{v\in V}\frac{d\theta_v}{2\pi},
 \qquad \theta\in[0,2\pi)^V.
\]
Write $r_k=I_k(\log\beta)/I_0(\log\beta)$, where $I_k$ is the
modified Bessel function of the first kind. Fix a reference orientation
of $E$, and let $\operatorname{div}j(v)$ be the outgoing minus incoming
sum of the signed edge values at $v$. The Fourier expansion
\cite[Section~10.35]{DLMFSpins} gives
\begin{equation}\label{spin:eq:xy-flow}
 \pi_{\mathbb Z}(j)=\frac1{Z_{\mathbb Z}}
 \one_{\{\operatorname{div}j=0\}}\prod_{e\in E}r_{|j_e|},
 \qquad j\in\mathbb Z^E,
 \qquad Z_{\rm XY}=I_0(\log\beta)^mZ_{\mathbb Z}.
\end{equation}
The spin-recovery maps used below are stated in
\cref{spin:lem:ising-recovery,spin:lem:xy-recovery}.

\paragraph{Potts flows and random clusters.}
Fix an orientation of the edges. For $q\ge3$, put
$r=(\beta-1)/(q+\beta-1)$. Expanding each edge factor in the characters
of $\mathbb Z_q$ gives the conserved flow law
\[
 \pi_q(\xi)\propto
 \one_{\{\operatorname{div}\xi=0\}}r^{|\operatorname{supp}\xi|},
 \qquad \xi\in\mathbb Z_q^E.
\]
Its partition function differs from $Z_{q,\beta}$ by the explicit
factor $q^n[(\beta+q-1)/q]^m$. The associated random-cluster law is
\[
 \pi_{\rm RC}(A)\propto(\beta-1)^{|A|}q^{k(A)},\qquad A\subseteq E,
\]
where $k(A)$ counts connected components, including isolated vertices.
Giving each component an independent uniform color in $[q]$ recovers
the Potts law \cite{ES88}. We prove the flow-to-cluster coupling,
including the version with prescribed divergence, in
\cref{lem:flow-completion}.

\subsection{Polymer models}\label{pol:sec:definitions}

For a polymer model as in \cref{pol:def:model}, with a finite or
countable type space, write
\[
 A_{\gamma\gamma'}=\one_{\{\gamma\not\sim\gamma'\}},
 \qquad W=\Diag(w),
 \qquad D(\gamma)=\sum_{\gamma'\not\sim\gamma}w_{\gamma'}.
\]
Thus $A$ includes the diagonal, and $D(\gamma)$ is the total activity
conflicting with $\gamma$. Spectral bounds on $AW$ refer to the
symmetric realization $W^{1/2}AW^{1/2}$.

For continuously marked types, activities form a finite measure
$\omega$. The same model has partition function
\[
 Z=\sum_{k\ge0}\frac1{k!}\int
 \one_{\{\gamma_1,\ldots,\gamma_k\text{ pairwise compatible}\}}
 \prod_{i=1}^k d\omega(\gamma_i).
\]
Sums of activities become integrals. In particular, $AW$ denotes the
self-adjoint conflict operator on $L^2(\omega)$,
\[
 (AWf)(\gamma)=\int_{\gamma'\not\sim\gamma}f(\gamma')\,d\omega(\gamma'),
 \qquad D(\gamma)=\int_{\gamma'\not\sim\gamma}d\omega(\gamma').
\]
Discrete models correspond to $\omega=\sum_\gamma w_\gamma\delta_\gamma$.

\subsection{Covariance bounds}\label{sec:covariance-preliminaries}
For a finite polymer law, $X_i$ denotes the indicator that coordinate
$i$ is occupied. A feasible positive pinning conditions a compatible
set of coordinates to be occupied; its residual law is the hard-core
law on the remaining feasible coordinates. We use covariance bounds
of the form $\Cov(X)\preceq C\Diag(\E X)$, where $\preceq$ is the
positive-semidefinite order. Such bounds after positive pinning and
decreasing all activities provide the spectral-stability input below.

\subsection{Computational model}\label{sec:computational-model}
The graph is given by indexed incidence lists. Work counts arithmetic
operations, integer rounding, comparisons, evaluation of elementary functions, and
generation of uniform and geometric random variables. For continuous
spins we also permit evaluation of Bessel ratios and generation of
Poisson and beta random variables. These are arithmetic-work
guarantees in a real-arithmetic model. Preprocessing, unsuccessful
proposal trials, data-structure operations, and recovery of all spins
are included in the stated bounds.

We first remove isolated vertices in $O(n+m)$ work and sample their
spins independently and uniformly in $O(n)$ work. If no edges remain,
this finishes the sampling task. In the three application sections,
$n$ and $m$ refer to the remaining graph, so $m\ge1$ and $n\le2m$.
The main theorems include the preprocessing and isolated-spin costs
on the original input. The case $\beta=1$ is independent sampling
and takes $O(n+m)$ work.

\section{Polymer Sampling and Exploration}\label{pol:sec:sampling}
We prove \cref{pol:lem:sampling} by comparing the insertion--deletion
chain with a finite random pool. The pool supplies the marginal lower
bounds needed by the low-degree log--Sobolev criterion of
\cite[Theorem~1.4]{MY26Matchings}; its covariance integral determines
the two mixing bounds. Section~\ref{pol:sec:overview} gives this
reduction and proves the pool comparisons. Section~\ref{pol:app:stability}
establishes the covariance bounds. Finally,
Section~\ref{pol:sec:certificates} gives exploration tools for verifying
the spectral and activity hypotheses in the applications.

\subsection{Random discretization and mixing}\label{pol:sec:overview}
\label{pol:app:finite-law}\label{pol:app:comparison}\label{pol:sec:auxiliary}

Use the polymer-model notation of \cref{pol:sec:definitions}.
Throughout the sampling argument, assume the hypotheses of
\cref{pol:lem:sampling}. We first give a proposal implementation of
its insertion--deletion chain.
\Needspace{7\baselineskip}
\phantomsection\label{def:ins-del-chain}
Starting from the empty family, the insertion--deletion chain $P$
performs the following update:
\begin{enumerate}[label=(\roman*)]
\item With probability $\frac1{4N}\sum_\gamma w_\gamma$, draw an
independent polymer with law $w_\gamma/\sum_{\gamma'}w_{\gamma'}$,
returning its full model representation, and insert it if compatible
with the current family.
\item With probability $1/4$, choose $j$ uniformly from $\{1,\ldots,N\}$
and delete the $j$th present polymer if the family has at least $j$ members.
\item Otherwise keep the current family.
\end{enumerate}
Each legal insertion of $\gamma$ has probability $w_\gamma/(4N)$,
and each present polymer is deleted with probability $1/(4N)$, so
$P$ is lazy and $\mu$-reversible. A trial returning $\gamma$ with
probability $w_\gamma/(4N)$ implements the same update without
evaluating $\sum_\gamma w_\gamma$.

The pool below is used to analyze this chain. Its size may be polynomial
in $N/\varepsilon$, but it need not be constructed by the algorithm.
We state the pool bounds and comparisons first, use them to prove the
sampling theorem, and then give their proofs.

Following the random-discretization approach of \cite{FGKKP25Random},
choose a sufficiently large constant $C$ and set
$h=C\log(N/\varepsilon)$ and $B=(N/\varepsilon)^C$.
Form a pool $\mathcal Q$ by drawing an independent
$\operatorname{Poisson}(Bw_\gamma)$ number of copies for each type
$\gamma$ with $D(\gamma)\le h$.
Each point is a separate coordinate of activity $1/B$, with conflicts
inherited from its type. Write $\pi_{\rm p}$ for the pool's hard-core law.
On the event $\mathcal G$ below, let $P_{\rm p}$ insert each feasible
copy with probability $1/(4NB)$ and delete each present copy with
probability $1/(4N)$. The spectral transfer uses matrix sampling
\cite[Theorem~6.4]{SW25Sampling}.

\begin{lemma}[Pool bounds]\label{pol:lem:finite-law}
The constant $C$ can be chosen, depending only on $\eta,\delta$, so
that the following event $\mathcal G$ has probability at least
$1-\varepsilon/100$:
\begin{enumerate}[label=(\roman*)]
\item $|\mathcal Q|\le2BN$ and the maximum closed conflict degree is
$O(Bh)$;
\item every feasible occupied marginal after positive pinning is at
least $B^{-1}e^{-O(h)}$;
\item the conflict spectrum satisfies
\[
 \lambda_{\min}(AW)\ge-1+\delta
 \ \Longrightarrow\
 \lambda_{\min}(A_{\rm p}/B)\ge-1+3\delta/4,
\]
\[
 \lambda_{\max}(AW)\le e-\delta
 \ \Longrightarrow\
 \lambda_{\max}(A_{\rm p}/B)\le e-3\delta/4.
\]
\end{enumerate}
\end{lemma}

All pool laws are projected onto polymer types. The next lemma compares
both their stationary distributions and their trajectories from the empty
family with those of $P$.

\begin{lemma}[Equilibrium and trajectory comparison]\label{pol:lem:comparison}
For either horizon $L$ in \cref{pol:lem:sampling}, the constants in the
pool construction can be chosen so that
\[
 \TV\!\left(\E[\pi_{\rm p}\mid\mathcal G],\mu\right)\le\varepsilon/4,
\]
and
\[
 \sup_{0\le j\le L}
 \TV\!\left(
   \E[\delta_{\varnothing}P_{\rm p}^{\,j}\mid\mathcal G],
   \delta_{\varnothing}P^j
 \right)\le\varepsilon/4.
\]
\end{lemma}

The remaining input is a covariance bound for the finite pool under
decreasing activities. For $0\le t<1$, let $\pi_{{\rm p},t}$ have
activities $(1-t)/B$. The following estimate is proved in
Section~\ref{pol:app:stability}.

\begin{lemma}[Spectral stability of the pool]\label{pol:lem:pool-stability}
On $\mathcal G$, there is a nonincreasing profile $C(t)$ such that,
after every feasible positive pinning $\tau$,
\[
 \Cov_{\pi_{{\rm p},t}^{\tau}}(X)
 \preceq C(t)\Diag\!\left(\E_{\pi_{{\rm p},t}^{\tau}}X\right),
\]
and
\[
 \int_0^1\frac{C(t)-1}{1-t}\,dt=
 \begin{cases}
 O_\delta(1),&\lambda_{\min}(AW)\ge-1+\delta,\\
 O_\delta(\sqrt h),&\lambda_{\max}(AW)\le e-\delta.
 \end{cases}
\]
\end{lemma}

The following specialization of the low-degree approximation criterion
converts this covariance integral and the pool's marginal lower bound
into a log--Sobolev inequality.

\begin{lemma}[Finite-model log--Sobolev input
{\cite[Theorem~1.4]{MY26Matchings}}]\label{pol:lem:external-lsi}
Let $\pi$ be a finite hard-core law on $M$ coordinates, with activities
at most one. Suppose every feasible occupied marginal after positive
pinning is at least $b>0$. Under activities multiplied by $1-t$,
assume that all such residual laws satisfy
\[
 \Cov(X)\preceq C(t)\Diag(\E X),\qquad 0\le t<1,
\]
for a nonincreasing $C(t)\ge1$, and put
$I=\int_0^1(C(t)-1)/(1-t)\,dt<\infty$.
For uniform-coordinate heat-bath dynamics $P_{\rm HB}$,
\[
 \Ent_\pi(f^2)\le O\!\left(Me^I\log\frac Mb\right)
                   \mathcal E_{P_{\rm HB}}(f,f).
\]
\end{lemma}

This is \cite[Theorem~1.4]{MY26Matchings} with $N=M$ and
$\kappa=1$: the hard-core support is downward closed, and each
feasible insertion ratio equals its activity, hence is at most one.
The occupied-marginal bound $b$, covariance normalization
$\Diag(\E X)$, and uniform-coordinate heat-bath kernel agree
with those in that theorem.

\begin{proof}[Proof of \cref{pol:lem:sampling}]
\phantomsection\label{proof:pol:lem:sampling}
Project all pool configurations onto polymer types. The triangle
inequality gives
\begin{equation}\label{pol:eq:tv-comparison}
\begin{aligned}
 \TV(\delta_{\varnothing}P^T,\mu)
 \le{}&\TV\!\left(\delta_{\varnothing}P^T,
       \E[\delta_{\varnothing}P_{\rm p}^{\,T}\mid\mathcal G]\right)\\
 &+\E\!\left[\TV(\delta_{\varnothing}P_{\rm p}^{\,T},\pi_{\rm p})
                       \mid\mathcal G\right]
 +\TV\!\left(\E[\pi_{\rm p}\mid\mathcal G],\mu\right).
\end{aligned}
\end{equation}
The trajectory and equilibrium terms are at most $\varepsilon/4$
each by \cref{pol:lem:comparison}. It remains to bound the middle
term for each pool in $\mathcal G$.

Write $I$ for the integral in \cref{pol:lem:pool-stability}.
The pool has coordinate activity $1/B\le1$ and, by
\cref{pol:lem:finite-law}, its occupied-marginal bound satisfies
$\log(|\mathcal Q|/b)=O(\log(N/\varepsilon))$.
The heat-bath and insertion--deletion update rates obey
\begin{equation}\label{pol:eq:update-scale}
 \mathcal E_{P_{\rm p}}
 =\frac{|\mathcal Q|(B+1)}{4NB}\mathcal E_{P_{\rm HB}}.
\end{equation}
Applying \cref{pol:lem:external-lsi} and this identity yields
\begin{equation}\label{pol:eq:pool-lsi}
 \Ent_{\pi_{\rm p}}(f^2)
 \le O\!\left(Ne^I\log\frac N\varepsilon\right)
           \mathcal E_{P_{\rm p}}(f,f).
\end{equation}
The pool coordinate count cancels from the mixing scale.

Since $P_{\rm p}$ holds with probability at least $1/4$, its smallest
eigenvalue is at least $-1/2$. Conditional Jensen and spectral calculus
give, for every density $g$,
\begin{equation}\label{pol:eq:entropy-step}
 \Ent_{\pi_{\rm p}}(g)-\Ent_{\pi_{\rm p}}(P_{\rm p}g)
 \ge\mathcal E_{P_{\rm p}^2}(\sqrt g,\sqrt g)
 \ge\tfrac12\mathcal E_{P_{\rm p}}(\sqrt g,\sqrt g).
\end{equation}
Thus relative entropy contracts by
$1-\Omega((Ne^I\log(N/\varepsilon))^{-1})$ per step.
The empty family's initial relative entropy is
$\log Z_{\rm p}\le|\mathcal Q|/B\le2N$.
After $T=O(Ne^I\log^2(N/\varepsilon))$ steps, Pinsker's inequality
\eqref{mc:eq:pinsker}
bounds the middle term in \eqref{pol:eq:tv-comparison} by
$\varepsilon/4$. The two bounds on $I$ give the two asserted horizons.
\end{proof}

\paragraph{From steps to work.}
The mixing lemma charges an update regardless of whether its candidate
is accepted. Each application must implement a trial with output
measure $d\omega/(4N)$ and account for failed trials and compatibility
tests. Starting from the empty family allows deletion work to be
charged to insertions. For a processing cost $\ell(\gamma)$, the
expected total cost of proposed candidates in $T$ steps is
$\frac{T}{4N}\int\ell(\gamma)\,d\omega(\gamma)$.
This is the interface used in the three applications below.

\paragraph{Proofs of the pool bounds and comparisons.}
We first verify the spectral and marginal properties of the discretization,
then compare its law and proposal stream with the original model.

\begin{proof}[Proof of \cref{pol:lem:finite-law}]
\phantomsection\label{proof:pol:lem:finite-law}
Restrict $\omega$ to $D(\gamma)\le h$, and denote the resulting measure
by $\widehat\omega$. Restriction compresses the conflict operator, so
it preserves both spectral hypotheses.

First suppose there are finitely many types, of weights $w_i$.
Replace type $i$ by $\lfloor Rw_i\rfloor$ identical copies and pad the
closed conflict matrix with zero rows to dimension $\lceil RN\rceil$.
After division by $R$, its nonzero spectrum converges to that of the
weighted conflict matrix. Retain each coordinate independently with
probability $B/R$. Theorem~6.4 of \cite{SW25Sampling}, applied with
normalized error $\delta/(8N)$ and failure probability
$\varepsilon/1000$, approximates its spectrum to additive error
$\delta/8$ after rescaling. The expected sample size required by that
theorem is polynomial in $N/\varepsilon$ and $\delta^{-1}$, independent
of $R$. Thus $B=(N/\varepsilon)^C$ suffices. Letting $R\to\infty$
gives independent Poisson copy counts and the scaled sampled matrix
$A_{\rm p}/B$. Using the remaining spectral slack gives (iii).

For marked types, approximate the bounded symmetric conflict kernel
in $L^2(\widehat\omega\otimes\widehat\omega)$ by symmetric
$\{0,1\}$ kernels constant on finite measurable partitions. Refine the
partitions and put self-conflict on each diagonal block; atoms are
retained in the limit. The operators converge in Hilbert--Schmidt
norm. Couple the pools by sampling the same types. The expected
number of differing off-diagonal entries is bounded by
$B^2\iint|A-A_k|\,d\widehat\omega\,d\widehat\omega$, which tends to
zero. The finite-type spectral conclusion therefore passes to the
marked model.

The pool size is Poisson with mean at most $BN$, so it is at most
$2BN$ except with exponentially small probability. Conditional on a
pool point of type $\gamma$, the number of other conflicting points
is Poisson with mean at most $Bh$. A Chernoff bound and the Mecke
formula \cite[Chapter~4]{LP18Poisson} bound the expected number of
points with closed degree exceeding $3Bh$ by
$BN e^{-cBh}$. Increasing $C$ makes these failures, together with the
spectral failure, at most $\varepsilon/100$.

Positive pinning leaves an induced conflict graph. At a feasible
coordinate of degree $d\le3Bh$, restricting its neighbors to be
absent gives
\[
 \pi_{\rm p}^{\tau}(x\in X)
 \ge\frac{1/B}{1/B+(1+1/B)^d}
 \ge B^{-1}e^{-O(h)}.
\]
Indeed, restoring these neighbors multiplies the partition function
by at most $(1+1/B)^d$. This proves (ii).
\end{proof}

For the comparison, we follow the output-law argument of
\cite[Theorem~5.1]{FGKKP25Random} and couple the proposal streams of
the original and pool chains.

\begin{proof}[Proof of \cref{pol:lem:comparison}]
Restrict $\omega$ to $D\le h$, obtaining $\widehat\omega$, with law
$\widehat\mu$ and partition function $\widehat Z$.

\emph{Equilibria.}
For this Poisson discretization, factorial moments give
\[
 \E Z_{\rm p}=\widehat Z,
 \qquad
 \E[(Z_{\rm p}/\widehat Z)\pi_{\rm p}]=\widehat\mu.
\]
They follow by integrating each compatible tuple in the partition
function; copies of an identical type cannot coexist. Adding one
copy increases $Z_{\rm p}$ by at most $Z_{\rm p}/B$. The Poisson
Poincar\'e inequality \cite[Theorem~18.7]{LP18Poisson} then gives
\[
 \Var Z_{\rm p}\le\frac NB\E Z_{\rm p}^2,
 \qquad
 \TV(\E\pi_{\rm p},\widehat\mu)
 \le\frac12\E|Z_{\rm p}/\widehat Z-1|
 \le\frac12\sqrt{\frac N{B-N}}.
\]
The exponential activity bound implies
$\omega(D>h)\le Ne^{-\eta h}$. The probability that a polymer from
this set is occupied is at most its total activity, so
$\TV(\widehat\mu,\mu)\le Ne^{-\eta h}$.
Conditioning the pool on $\mathcal G$ costs at most
$\Prb(\mathcal G^c)$ in TV.

\emph{Discrete trajectories.}
Conditional on its size $K$, the pool consists of independent types
with law $\widehat\omega/\widehat\omega(\mathcal P)$. At each step its insertion attempt
probability is $K/(4NB)$; conditional on attempting, it chooses a
uniform copy. The reference chain attempts with probability $\widehat\omega(\mathcal P)/(4N)$.
Use the same uniform variable for these choices and the same deletion
choice in both chains. The probability of any differing attempt decision in $L$ steps is at most
\[
 \frac L{4N}\E|K/B-\widehat\omega(\mathcal P)|
 \le\frac L{4N}\sqrt{N/B}.
\]
Reveal a copy's type on its first use. Until a copy is chosen twice,
these types form independent reference candidates. For any two steps,
conditional on $K$, the probability that both choose the same copy
is $K/(16N^2B^2)$. Hence the probability of any such collision is at
most $\binom L2/(16NB)$.
Identical candidates give identical compatibility decisions and
projected states. On the event $K>2BN$ define the auxiliary kernel
arbitrarily; this event is contained in $\mathcal G^c$.

Removing types with $D>h$ changes a proposal with probability at most
$\omega(D>h)/(4N)$ per step. Combining this with the preceding
coupling and conditioning on $\mathcal G$ yields
\[
 \sup_{j\le L}\TV\!\left(
 \E[\delta_\varnothing P_{\rm p}^{\,j}\mid\mathcal G],
 \delta_\varnothing P^j\right)
 \le \frac L4 e^{-\eta h}
 +\frac L{4N}\sqrt{\frac NB}
 +\frac{L^2}{32NB}+2\Prb(\mathcal G^c).
\]
For either stated horizon, $\log(L/N)=O_\delta(\sqrt h+\log h)$.
Choosing $C$ sufficiently large makes this bound and the equilibrium
bound at most $\varepsilon/4$.
\end{proof}

\subsection{Covariance bounds from the conflict spectrum}
\label{pol:app:stability}
We prove \cref{pol:lem:pool-stability}. The lower-spectrum case follows
from \cite[Theorem~1.14]{CCCYZ25}. For the upper-spectrum case, we first
establish a covariance estimate for a finite hard-core model, then apply
it to the pool. Here $A^\circ$ excludes self-conflicts.

\begin{lemma}[Covariance below weighted radius $e$]\label{pol:lem:local-covariance}
Let a finite hard-core model have reference activities $w_i>0$.
Suppose $u_i>0$ and $c<e$ satisfy
$\sum_j A^\circ_{ij}w_ju_j\le cu_i$. After any feasible pinning,
give the remaining coordinates activities $w'_i\le a w_i$, where
$0<a\le1$, and write $p_i$ for their occupied marginals. On the
nondeterministic coordinates,
\[
 \Cov(X)\preceq
 \Diag\!\left(p_i\left[1+
 \frac{a\sqrt{c\sum_j A^\circ_{ij}w_j}}
      {1-\sqrt{ac/e}}\right]\right).
\]
\end{lemma}

\begin{proof}[Proof of \cref{pol:lem:local-covariance}]
\phantomsection\label{proof:pol:lem:local-covariance}
We use the occupation-odds recursion with potential
$\varphi(R)=2\operatorname{arsinh}\sqrt R$.
Expand the exact deletion recursion into a computation tree:
at a node $v$ representing coordinate $i(v)$,
\[
 R_v=w'_{i(v)}\prod_{z\text{ child of }v}(1+R_z)^{-1}.
\]
The children are the successive neighbor-deletion subinstances in a
fixed order; their labels are distinct neighbors of $i(v)$.
Write $q_v=R_v/(1+R_v)$. In this potential the absolute derivative
across an edge $vz$ is $\sqrt{q_vq_z}$.

Use weights $\zeta_i=\sqrt{w_i u_i}$. With $s_v=\sum_z q_z$, the recursion gives
$q_v\le a w_{i(v)}e^{-s_v}$. Cauchy--Schwarz and the witness bound yield
\[
 \sum_z\sqrt{q_vq_z}\,\zeta_{i(z)}
 \le\sqrt{q_vs_v}\sqrt{\sum_z w_{i(z)}u_{i(z)}}
 \le\sqrt{ac\,s_ve^{-s_v}}\,\zeta_{i(v)}
 \le\sqrt{ac/e}\,\zeta_{i(v)}.
\]
This contracts the derivatives below the root.

To normalize by the actual marginals, put
$\Sigma=\Diag(p)^{-1/2}\Cov(X)\Diag(p)^{-1/2}$ and
$g_i=\sqrt{p_i u_i}$. Differentiating the recursion rooted at $i$
with respect to $\log w'_j$ gives, for $i\ne j$,
\[
 |\Sigma_{ij}|\le(1-p_i)
 \sum_{v:i(v)=j}
 \left(\prod_{xy\in\operatorname{path}(i,v)}\sqrt{q_xq_y}\right)
 \sqrt{q_v/p_j}.
\]
Each occurrence of $j$ contributes one source derivative to its
log-odds. The endpoint factor obeys
$\sqrt{q_v/p_j}\,g_j\le\sqrt a\,\zeta_j$.
At the root, retain $q_i=p_i$ and sum the deeper levels geometrically.
Since the root child sum is at most
$a\sum_j A^\circ_{ij}w_j$, this gives
\[
 \frac{\sum_j|\Sigma_{ij}|g_j}{g_i}
 \le(1-p_i)\left(1+
       \frac{\sqrt{ac\,s_i}}{1-\sqrt{ac/e}}\right)
 \le1+\frac{a\sqrt{c\sum_j A^\circ_{ij}w_j}}{1-\sqrt{ac/e}}.
\]
For a symmetric matrix, the weighted row bound
$\sum_j|\Sigma_{ij}|g_j\le d_i g_i$ implies $\Sigma\preceq\Diag(d)$:
apply $2|x_ix_j|\le x_i^2g_j/g_i+x_j^2g_i/g_j$ to its quadratic
form. Conjugating by $\Diag(\sqrt p)$ proves the claim. Pinning
leaves an induced residual graph, preserving the witness inequality.
\end{proof}

\begin{proof}[Proof of \cref{pol:lem:pool-stability}]
\phantomsection\label{proof:pol:lem:pool-stability}
On $\mathcal G$, the pool has closed degree at most $3Bh$.
In the lower-spectrum case, subtracting the diagonal $I/B$ from
$A_{\rm p}/B$ gives
$A^\circ_{\rm p}/B\succeq-(1-\delta/2)I$.
After any feasible positive pinning, let $R$ be the remaining
available coordinates. The pairwise dependency matrix of
\cite[Definition~1.6]{CCCYZ25} is
$M_R=-A^\circ_{\rm p}[R]$, and every feasible insertion ratio is $1/B$.
The principal-submatrix bound gives
$M_R\preceq B(1-\delta/2)I\preceq I+B(1-\delta/2)I$,
which verifies the hypothesis of
\cite[Theorem~1.14]{CCCYZ25} with margin $\delta/2$ on every residual
graph. Applying its scalar tilt $1-t$ gives the profile
\[
 C(t)=\frac1{1-(1-\delta/2)(1-t)},
 \qquad \int_0^1\frac{C(t)-1}{1-t}\,dt=\log(2/\delta).
\]
In the upper-spectrum case, put $c=e-\delta/2$. The positive resolvent
of $cI-A^\circ_{\rm p}/B$ supplies a positive witness with weighted
radius at most $c$. Apply \cref{pol:lem:local-covariance} with
$a=1-t$ and weights $1/B$. We may take
\[
 C(t)=1+\frac{(1-t)\sqrt{3ch}}{1-\sqrt{(1-t)c/e}},
 \qquad \int_0^1\frac{C(t)-1}{1-t}\,dt=O_\delta(\sqrt h).
\]
\end{proof}

\subsection{Exploration certificates}\label{pol:sec:certificates}
The following two tools verify the hypotheses of the sampling theorem
in the applications. First-hit probabilities give an upper spectral
bound. Joint completion events give a lower spectral bound and the
exponential activity estimate.

\paragraph{First-hit bounds.}
We seek positive weights $h_\gamma$ such that
\[
 \sum_{\gamma'\not\sim\gamma}w_{\gamma'}h_{\gamma'}
 \le\kappa h_\gamma.
\]
This bounds the total weighted influence of polymers that conflict
with $\gamma$, and implies $\lambda_{\max}(AW)\le\kappa$.
The next lemma constructs these weights from a random exploration:
$h_\gamma$ is the sum, over starting states, of the probabilities of
hitting $\gamma$. The key requirement is that, after hitting a support
$F$, the exploration completes it with probability at least
$w_F/\kappa$. If an exploration completes at most one support, and
completing a support also hits every support intersecting it, the
desired inequality follows by summing these mutually exclusive events.
In the applications, this completion probability is checked at the
first visit to $F$. The same explorations also generate insertion
proposals after correcting their output laws.

\begin{lemma}[First-hit bound]\label{pol:lem:first-hit}
Give each polymer a nonempty support $F_\gamma$ in a finite ground set,
so that incompatible polymers have intersecting supports. Let
$\mathcal F$ be the distinct supports and let $w_F$ be the total
activity of types with support $F$, integrating over marks when present.
For each start $s$ in a finite set $\mathcal S$, consider an exploration.
Write $\Prb_s$ for its law when started at $s$, and let $H_F$ be the
event that it visits at least one element of $F$. Specify an event
$E_F$ that the exploration completes $F$. Assume that each support
has a starting state from which it is hit with probability one.
For a constant $\kappa>0$, suppose that, for every start $s$:
\begin{enumerate}[label=(\roman*)]
\item \emph{At most one completion:} the events $E_F$, $F\in\mathcal F$,
are mutually exclusive.
\item \emph{Completion hits intersecting supports:}
$E_F\subseteq H_{F'}$ whenever $F\cap F'\ne\varnothing$.
In particular, completing $F$ entails hitting $F$.
\item \emph{Completion after a hit:} for every $F$ with
$\Prb_s(H_F)>0$,
\[
 \Prb_s(E_F\mid H_F)\ge w_F/\kappa,
 \quad\text{equivalently}\quad
 w_F\Prb_s(H_F)\le\kappa\Prb_s(E_F).
\]
\end{enumerate}
Then $h_\gamma=\sum_{s\in\mathcal S}\Prb_s(H_{F_\gamma})$ satisfies
\begin{equation}\label{pol:eq:first-hit}
 AWh\le\kappa h,\qquad
 1\le h_\gamma\le|\mathcal S|,\qquad
 \sum_\gamma w_\gamma h_\gamma\le\kappa|\mathcal S|.
\end{equation}
In particular, $\lambda_{\max}(AW)\le\kappa$ and
$\sum_\gamma w_\gamma\le\kappa|\mathcal S|$.
\end{lemma}
\begin{proof}
Fix a polymer $\gamma$ and a start $s$. The completion bound,
mutual exclusion, and the hitting implication give
\[
 \sum_{F:F\cap F_\gamma\ne\varnothing}w_F\Prb_s(H_F)
 \le\kappa\sum_{F:F\cap F_\gamma\ne\varnothing}\Prb_s(E_F)
 \le\kappa\Prb_s(H_{F_\gamma}).
\]
Summing over starts bounds $AWh$ by $\kappa h$, since every conflict
is included among these support intersections. Summing the completion
bound over all supports gives $\sum_\gamma w_\gamma h_\gamma\le
\kappa|\mathcal S|$. The bounds on $h$ follow from the certain hitting
start and its definition. The weighted Schur bound gives the spectral
conclusion, and $h\ge1$ gives the total activity bound.
\end{proof}

\begin{example}[A nonbacktracking walk completing a cycle]
\label{pol:ex:first-hit-cycle}
The statement of \cref{pol:lem:first-hit} is abstract. We illustrate
its notation with the cycle exploration used in \cref{spin:sec:ising},
in the special case of a graph $H$ of maximum degree three with
constant edge weight $0<r\le1/2$. A polymer is a cycle $C$, its
support is the set $F_C=E(C)$ of undirected edges, and its activity
is $w_C=r^{|C|}$. Two cycles conflict when they share a vertex;
at maximum degree three they then also share an edge.

The starting states are directed edges, $\mathcal S=\vec E(H)$.
From $a=(u,v)$, move to each available successor $(v,w)$ with $w\ne u$
with probability $r$, and stop with the remaining probability.
There are at most two successors, so these probabilities are valid.
Also stop as soon as a directed edge is visited for the second time.
The visited set $\mathcal R$ records the underlying undirected edges,
including the starting edge.

\begin{center}
\begin{tikzpicture}[>=Stealth,scale=0.95,
 vertex/.style={circle,fill=black,inner sep=1.7pt},
 route/.style={->,very thick,blue!65!black}]
 \coordinate (u) at (0,0);
 \coordinate (v) at (2.5,0);
 \coordinate (w) at (2.5,2);
 \coordinate (x) at (0,2);
 \coordinate (q) at (4.2,1);
 \draw[thick,dashed,orange!80!black] (v)--(q)--(w);
 \draw[route] (-2,-0.8)--node[below,sloped] {$s$}(-1,-0.4);
 \draw[route] (-1,-0.4)--(u);
 \draw[route] (u)--node[above] {$a_0$: first hit}(v);
 \draw[route] (v)--node[left] {$a_1$}(w);
 \draw[route] (w)--node[above] {$a_2$}(x);
 \draw[route] (x)--node[left] {$a_3$}(u);
 \draw[->,blue!65!black,thick] (0.15,-0.15) to[bend right=28]
 node[below] {return to $a_0$; stop} (2.35,-0.15);
 \node[blue!65!black] at (1.15,1) {$C$};
 \node[orange!80!black] at (3.15,1) {$C'$};
 \node[align=left,anchor=west,font=\small] at (5,1.5)
 {$F_C=\{uv,vw,wx,xu\}$\\[4pt]
 $F_C\cap F_{C'}=\{vw\}$\\[4pt]
 $w_C=r^4$};
 \foreach \p/\pos in {u/below left,v/below right,w/above right,x/above left,q/right}
  \node[vertex,label=\pos:{$\p$}] at (\p) {};
\end{tikzpicture}

\small The blue arrows show a successful exploration. The curved arrow
denotes a second traversal of $uv$, not an additional graph edge.
\end{center}

Here $H_C$ means that the walk visits an edge of $C$, in either
direction. Let $\tau_C$ be its first such visit. In the picture the
state at $\tau_C$ is $a_0=(u,v)$. The event $E_C$ means that the next
four transitions are
\[
 a_0\longrightarrow a_1\longrightarrow a_2
 \longrightarrow a_3\longrightarrow a_0.
\]
Thus hitting $C$ is not yet completing it: the walk could stop or
leave $C$ at $v$ or $w$. Conditional on any history up to the first
hit, the displayed completion has probability $r^4=w_C$.
No edge of $C$ was visited before $\tau_C$, so the last transition
is the first repeated directed edge. For any cycle, the same argument
gives $\Prb_s(E_C\mid H_C)=r^{|C|}=w_C$ whenever the conditioning
event has positive probability.

The three hypotheses now have concrete meanings. The terminal circuit
determines the completed cycle, so different completion events are
mutually exclusive. Completing $C$ traverses every edge of $C$, hence
hits every intersecting support; for example, it hits $C'$ through
$vw$ in the picture. Finally, the completion probability is exactly
$w_C$, so $\kappa=1$. Starting on $C$ guarantees a hit. Consequently,
\[
 h_C=\sum_{s\in\vec E(H)}\Prb_s(\text{the walk hits an edge of }C)
 \quad\text{satisfies}\quad
 \sum_{C'\not\sim C}w_{C'}h_{C'}\le h_C.
\]
This is the spectral certificate $AWh\le h$ needed in
\cref{spin:sec:ising}. There, nonuniform edge weights require a
normalization of the transition probabilities, but their product
around a cycle is still $w_C$, so the same first-hit argument applies.
\end{example}

\paragraph{Independent-event bounds.}
The next lemma supplies a lower spectral bound and an exponential
activity estimate when conflicts can be encoded by mutually exclusive
events. Its hypotheses are stronger than the first-hit conditions and
will be verified separately for the Ising and Potts representations.

\begin{lemma}[Event bounds]\label{spin:lem:event-bounds}
Let events $E_i$ have positive probabilities $p_i$ and a conflict
graph such that every joint intersection has probability
$\prod_{i\in T}p_i$ for independent $T$, and zero otherwise. Suppose
at most $K$ events occur simultaneously. For every partition $L\cup R$,
\[
 \left\|\bigl(\sqrt{p_ip_j}\,\one_{\{i\in N(j)\}}\bigr)_{i\in L,j\in R}\right\|
 \le\frac56,
 \qquad
 \sum_i p_i\exp\!\left(\frac12\sum_{j\in N[i]}p_j\right)
 \le2\sqrt e\,K.
\]
Here $N(i)$ and $N[i]$ are the open and closed conflict neighborhoods.
\end{lemma}

\begin{proof}
For the spectral bound, delete conflicts within each part. This
preserves realizability of the joint probabilities. Indeed, all
avoidance polynomials
$V(U)=\sum_{T\subseteq U\text{ independent}}(-1)^{|T|}\prod_{i\in T}p_i$
are nonnegative; deleting an edge $ab$ adds
$p_ap_bV(U\setminus(N[a]\cup N[b]))$ when $a,b\in U$.
The exact-occurrence probabilities
$\prod_{i\in I}p_i\,V(N[I]^c)$ realize the new graph.

Assume first $0<p_i<1$. Conditional on no event in $L$, set
\[
 s_j=-\sum_{i\in N(j)}\log(1-p_i),\qquad
 h_{jk}=-\sum_{i\in N(j)\cap N(k)}\log(1-p_i),\qquad
 r_j=p_je^{s_j}\quad(j,k\in R).
\]
The conditional means and distinct joint moments are $r_j$ and
$r_jr_ke^{-h_{jk}}$. Positivity of their covariance gives
$\Diag(1-p_j)\succeq T$, where
$T_{jk}=\sqrt{r_jr_k}(1-e^{-h_{jk}})$, including
$T_{jj}=r_j-p_j$. Write
$M_{ij}=\sqrt{p_ip_j}\one_{\{i\in N(j)\}}$ for $i\in L,j\in R$.
Using $1-e^{-h}\ge he^{-h/2}$ and $h_{jk}\le(s_j+s_k)/2$ gives the
entrywise bound
\[
 T\ge\Diag(e^{s/4})M^{\mathsf T}M\Diag(e^{s/4}).
\]
Perron monotonicity therefore yields $\|M\Diag(e^{s/4})\|\le1$.
For nonnegative unit singular vectors $My=\sigma x$ and
$M^{\mathsf T}x=\sigma y$, Jensen gives
$\sum_j y_j^2s_j\le-4\log\sigma$. Exchanging the parts gives the
corresponding bound on $L$. Cauchy--Schwarz now yields
\[
 \sigma^2\le
 \left(\sum_i x_i^2\sum_{j\in N(i)}p_j\right)
 \left(\sum_j y_j^2\sum_{i\in N(j)}p_i\right)
 \le(-4\log\sigma)^2.
\]
Thus $\sigma\le-4\log\sigma$, which implies $\sigma<5/6$.

For the exponential bound, use the original graph and initially
assume its full avoidance probability is positive. First condition on
the absence of $L\setminus N(i)$; joint factorization leaves the
probability of $E_i$ equal to $p_i$. Then expose absences of its
neighbors in $L$ successively. This gives
\[
 \Prb(E_i\mid\text{no event in }L)
 \ge p_i\exp\!\left(\sum_{j\in L\cap N(i)}p_j\right)
 \qquad(i\notin L).
\]
To justify each step, inclusion--exclusion gives
$\Prb(E_j\mid\text{no event in }U)=p_jV(U\setminus N(j))/V(U)\ge p_j$;
conditioning on a conflicting event's absence divides the current
probability by at most $1-p_j$. Summing over $i\notin L$ is at most $K$.
Choose $L$ by independent fair coins and use Jensen to obtain
$\sum_i p_i e^{\sum_{j\in N(i)}p_j/2}\le2K$.
Including the diagonal costs at most $\sqrt e$.
Independent thinning of all events, followed by a limit, removes
the positivity assumptions.
\end{proof}

\section{Critical Ising Sampling}\label{spin:sec:ising}

Consider the even-subgraph law \eqref{spin:eq:ising-flow} on a graph
$G$ of maximum degree $\Delta\ge3$, with uniform edge weight $r$ and
$(\Delta-1)r\le1$. If $G$ has no cycles, its only even subgraph is
the empty edge set: every nonempty subgraph of a forest has a vertex
of degree one. In this case no even-subgraph sampling is needed;
we apply spin recovery directly to the empty edge set on $G$.
Henceforth assume that $G$ contains a cycle. We first reduce to
maximum degree three, where the nonempty connected components of
an even subgraph are cycles.

\subsection{Degree reduction and cycle representation}\label{spin:app:ising}
The reduction preserves the nonbacktracking spectral bound
$\rho(B_t)\le1$ needed for the exploration. For a weighted graph $H$,
write $B_t$ for its nonbacktracking matrix.
For a directed edge $a$, let $e(a)$ be its physical edge and $\bar a$
its reversal. Define
\[
 (B_t)_{ab}=t_{e(b)}\one_{\{\operatorname{head}(a)=\operatorname{tail}(b),\ b\ne\bar a\}}.
\]

\begin{proposition}[Reduction to maximum degree three]\label{spin:prop:subcubic}
Let $G$ have maximum degree at most $\Delta\ge3$ and uniform flow
weight $r>0$, with $(\Delta-1)r\le1$. There is a graph $H$ of maximum
degree at most three, with at most $3m$ edges and weights
$t_e\in\{r,1\}$, whose even subgraphs are in a weight-preserving
bijection with those of $G$. Its nonbacktracking matrix satisfies
$\rho(B_t)\le1$. If $K$ retains only transitions within strongly
connected components of $B_t$, a positive vector $z$ with $Kz\le z$
can be constructed explicitly, giving the transition probabilities
in \eqref{spin:eq:ising-successor}.
\end{proposition}

\paragraph{The degree reduction.}
For \cref{spin:prop:subcubic}, repeatedly delete vertices of degree
zero or one; the remaining graph is the $2$-core. No deleted edge
can belong to an even subgraph. Replace each remaining degree-$d$
vertex by a path of $d$ ports. Attach its incident
edges to the ports in a fixed order. Original edges have weight $r$;
path edges have weight one. The resulting graph is $H$.
Set $z_a=1$ on original directed edges.
On a directed path edge, set $z_a=r$ times the number of ports ahead
in its direction.

\begin{proof}[Proof of \cref{spin:prop:subcubic}]
\phantomsection\label{proof:spin:prop:subcubic}
Pruned edges carry zero flow. At a degree-$d$ vertex, the labels on
the $d-1$ path edges are the successive parities of its incident
original-edge labels. The final parity is zero precisely when the
original degree is even. This gives a bijection of even subgraphs;
the added edges have weight one, so it preserves weights. There are
at most $2m$ ports and fewer than $2m$ added edges, hence
$|E(H)|\le3m$ and the maximum degree is three.

For the stated vector $z$, the outgoing weighted sum at a path-edge
arrival equals $z_a$: its available ports split into the current
exit and the ports farther along the path. At an original-edge
arrival the sum is $r(d-1)\le1=z_a$. Thus $B_tz\le z$, proving
$\rho(B_t)\le1$. Deleting transitions between strongly connected
components preserves this inequality and every cycle product, giving
the normalization in \eqref{spin:eq:ising-successor}.
\end{proof}

Every nonempty connected component of an even subgraph of $H$ is a
cycle. Define a polymer to be the edge set
of a cycle $C$, with
\[
 w_C=\prod_{e\in C}t_e,
 \qquad C\not\sim C'\ \Longleftrightarrow\ V(C)\cap V(C')\ne\varnothing.
\]
Two parallel edges may form a cycle. Taking the union of a compatible
cycle family gives exactly the even-subgraph law
\eqref{spin:eq:ising-flow}.

\subsection{Exploration and first-hit bound}
By \cref{spin:prop:subcubic}, $\rho(B_t)\le1$. Retain transitions
within the strongly connected components of $B_t$, obtaining $K$.
Choose $z>0$ with $Kz\le z$, and set
\begin{equation}\label{spin:eq:ising-successor}
 P_H(a,b)=K_{ab}\frac{z_b}{z_a}.
\end{equation}
The normalization preserves the product $w_C$ around either orientation
of each cycle. Starting at a directed edge $a$, the exploration moves
to $b$ with probability $P_H(a,b)$ and stops with probability
$1-\sum_bP_H(a,b)$. It also stops immediately upon reaching a directed
edge already visited. 

\begin{proposition}[The cycle first-hit bound]\label{spin:prop:ising-hit}
For the cycle model on $H$, if $\rho(B_t)\le1$, there is $h>0$ with
\[
 AWh\le h,\qquad 1\le h_C\le2|E(H)|,\qquad
 \sum_Cw_Ch_C\le2|E(H)|.
\]
\end{proposition}

\begin{proof}
Use all directed edges of $H$ as starts. Let $H_C$ be the event of
visiting either orientation of an edge of $C$, including at time zero.
Fix $C$ and let $\tau_C$ be the first time the exploration visits
either orientation of one of its edges. On $H_C$, the state at
$\tau_C$ fixes an orientation of $C$. Define $E_C$ to be the event
that the next $|C|$ transitions follow that oriented circuit and
return to the state at $\tau_C$. No edge of $C$ was visited before the hit,
so the return is the first repeated directed edge. Conditional on
any history up to $\tau_C$, the probability of this completion is
\[
 \prod_{(a,b)\text{ around }C}P_H(a,b)
 =\prod_{e\in C}t_e=w_C.
\]
The terminal circuit determines $C$, so the events $E_C$ are mutually
exclusive. Two cycles sharing a vertex in a graph of maximum degree
three share an edge; completing one therefore hits the other.
A start on $C$ makes $H_C$ certain. Apply
\cref{pol:lem:first-hit} with edge supports, $\kappa=1$, and
$\mathcal S=\vec E(H)$.
\end{proof}

\subsection{Spectral and activity bounds}
The first-hit bound gives $\lambda_{\max}(AW)\le1$, and hence
$\lambda_{\min}(AW)\ge-1$, but the near-linear branch of
\cref{pol:lem:sampling} requires a fixed strict margin.
We obtain that margin from an auxiliary successor map whose cycle
events either factor or are mutually exclusive. This map is used
for the spectral estimate; proposals still use the exploration above.

\begin{lemma}[Cycle spectrum and activity bounds]\label{spin:lem:ising-structure}
For the cycle representation supplied by \cref{spin:prop:subcubic},
\begin{equation}\label{spin:eq:ising-inputs}
 \lambda_{\min}(AW)\ge-5/6,\qquad
 \sum_Cw_Ce^{D(C)/2}\le2\sqrt e\,|E(H)|,\qquad
 \sum_C|C|w_C\le|E(H)|.
\end{equation}
\end{lemma}

\begin{proof}[Proof of \cref{spin:lem:ising-structure}]
\phantomsection\label{proof:spin:lem:ising-structure}
We choose incidence values to satisfy two requirements: the local
transition probabilities must sum to at most one, and the probabilities
of the two orientations of a cycle must have product $w_C^2$.
At port $i$ of an original degree-$d$ vertex, assign incidence values
\[
 x_{\rm original}=r,\qquad
 x_{\rm left}=\frac{i-1}{d-i+1},\qquad
 x_{\rm right}=\frac{d-i}{i},\qquad b_a=\frac{x_a}{1+x_a},
\]
omitting absent incidences. Their $b_a$ sum is
$r/(1+r)+(d-1)/d\le1$. Define the auxiliary successor kernel by
\begin{equation}\label{spin:eq:ising-event-kernel}
 Q(a,c)=\frac{b_c}{1-b_{\bar a}}\,
 \one_{\{\operatorname{head}(a)=\operatorname{tail}(c),\ c\ne\bar a\}}.
\end{equation}
Its row sums are at most one, including when $(\Delta-1)r=1$.
Independently for each directed edge $a$, draw a successor $Y_a$
with probabilities $Q(a,c)$ and no successor with the remaining
probability. For an oriented cycle
$(C,\sigma)=(a_1,\ldots,a_\ell)$, with $a_{\ell+1}=a_1$, put
\[
 E_{C,\sigma}=\bigcap_{i=1}^{\ell}\{Y_{a_i}=a_{i+1}\},
 \qquad
 d_C^\sigma=\Prb(E_{C,\sigma})
           =\prod_{i=1}^{\ell}Q(a_i,a_{i+1}).
\]
The identity
\[
 \frac{b_c}{1-b_{\bar a}}\frac{b_{\bar a}}{1-b_c}=x_{\bar a}x_c
\]
gives $d_C^+d_C^-=w_C^2$: opposite incidences on a path edge have
product one, and on an original edge product $r^2$.

Distinct oriented cycles sharing a directed edge are mutually
exclusive; those with disjoint directed-edge sets depend on
disjoint successor variables. At most $2|E(H)|$ events can occur
simultaneously. Two conflicting unoriented cycles share an edge,
so their oriented conflict graph contains a pair of conflicts and
its reversed pair.

Put $S=W^{1/2}AW^{1/2}$. For a real vector $z$, place both
orientations of $C$ on the side determined by the sign of $z_C$.
The two reversed coefficients have sum at least
$2\sqrt{w_Cw_D}$ by AM--GM. The rectangular bound of
\cref{spin:lem:event-bounds} therefore gives
\[
 \sum_{C:z_C>0}\sum_{D:z_D<0}
       S_{CD}|z_Cz_D|
 \le\frac56\|z_+\|\|z_-\|.
\]
All within-side terms are nonnegative, so
$z^{\mathsf T}Sz\ge-(5/6)\|z\|^2$.

Let $D^{\rm dir}_{C,\sigma}$ be the sum of probabilities of oriented
events conflicting with $(C,\sigma)$, including itself. Reversed
conflicts and AM--GM give
$D(C)\le(D^{\rm dir}_{C,+}+D^{\rm dir}_{C,-})/2$.
Applying AM--GM once more and then the exponential event bound gives
\[
 \sum_C w_Ce^{D(C)/2}
 \le\frac12\sum_{C,\sigma}d_C^\sigma e^{D^{\rm dir}_{C,\sigma}/2}
 \le2\sqrt e\,|E(H)|.
\]
Finally, start the normalized exploration
\eqref{spin:eq:ising-successor} from a uniform directed edge and accept
only a simple cycle returning to that starting state. A cycle of
length $\ell$ has $2\ell$ directed roots, each of probability
$w_C/(2|E(H)|)$. Its output probability is
$\ell w_C/|E(H)|$, proving $\sum_C|C|w_C\le|E(H)|$.
\end{proof}

\subsection{Exact proposals, spin recovery, and work}
\begin{lemma}[Exact cycle proposals]\label{spin:lem:ising-oracle}
There is a trial based on \eqref{spin:eq:ising-successor} whose output is
either a cycle $C$ or no candidate, with
\[
 \Prb(\text{output }C)=w_C/|E(H)|.
\]
Its expected exploration length is $O(\log n)$, including unsuccessful
trials. Running the trial with probability $|E(H)|/(4N)$ implements
the insertion proposal of $P$.
\end{lemma}

\begin{proof}[Proof of \cref{spin:lem:ising-oracle}]
\phantomsection\label{proof:spin:lem:ising-oracle}
Use the last exploration in the preceding proof, with an independent
horizon having survival probability $1/\ell$ at transition $\ell$.
After querying transition $\ell$, continue to transition $\ell+1$
with probability $\ell/(\ell+1)$ if the exploration has not stopped.
Retain a result only when its first repeated directed edge is its
starting state and the resulting physical cycle is simple.
The horizon multiplies the preceding output probability by
$1/|C|$, giving $w_C/|E(H)|$.

There are at most $2|E(H)|$ distinct directed states before repetition,
so the expected number of queried transitions is at most
$\sum_{\ell\le2|E(H)|}1/\ell=O(\log n)$.
Recording the path, recognizing a simple cycle, and returning its
edge list take work proportional to the exploration length. Running the trial with probability $|E(H)|/(4N)$ gives the insertion
measure of $P$.
\end{proof}

\paragraph{Spin recovery and work.}

\begin{lemma}[Recovery of Ising spins]\label{spin:lem:ising-recovery}
Given the even-subgraph law \eqref{spin:eq:ising-flow}, add each missing
edge $e$ independently with probability $t_e$, then assign an independent
fair sign to each connected component. The resulting spins have law
$\mu_{\rm Ising}$. The map contracts TV distance and takes $O(m)$ work.
\end{lemma}

\begin{proof}[Proof of \cref{spin:lem:ising-recovery}]
\phantomsection\label{proof:spin:lem:ising-recovery}
The completion theorem of \cite[Theorem~3.5]{GrimmettJanson09}
maps the even-subgraph law to FK-Ising by adding each missing edge
with probability $t_e$. Edwards--Sokal cluster coloring \cite{ES88}
then gives the Ising law. One edge scan and one component search take
$O(m)$ work.
\end{proof}

\begin{proposition}[Ising work]\label{spin:prop:ising-work}
Taking $N=10m$ in \cref{pol:lem:sampling}, the cycle chain from
$\varnothing$ can be simulated for $t$ steps in $O(m+t\log n)$ expected
work. Spin recovery costs $O(m)$, giving $O(m\log^3(n/\varepsilon))$
at the first mixing horizon in \cref{pol:lem:sampling}.
\end{proposition}

\begin{proof}[Proof of \cref{spin:prop:ising-work}]
\phantomsection\label{proof:spin:prop:ising-work}
Construct the port graph explicitly in $O(m)$ work. It has at most
$3m$ edges and degree at most three. Precompute its successor rows
and strongly connected components. A trial of
\cref{spin:lem:ising-oracle} queries each row in constant work, so its
expected cost is $O(\log n)$.

Store the owner of every occupied vertex of $H$. Checking or inserting
a cycle costs its length; store its edge list for deletion. Charge deletion work to insertion. Touched-vertex lists clear the temporary
exploration data without scanning $H$ after each trial. Thus $t$ chain
steps cost $O(m+t\log n)$ in expectation, including all trials,
compatibility tests and updates. Projection to original edges and
\cref{spin:lem:ising-recovery} take $O(m)$ work, since $n\le2m$.
\end{proof}

\begin{proof}[Proof of \cref{spin:thm:ising}]
\phantomsection\label{proof:spin:thm:ising}
Apply \cref{spin:prop:subcubic} with $r=(\beta-1)/(\beta+1)$.
By \cref{spin:lem:ising-structure}, \cref{pol:lem:sampling} applies with
$N=10m$, $\eta=1/2$, and a fixed lower spectral margin.
Its $O(m\log^2(n/\varepsilon))$ steps cost
$O(m\log^3(n/\varepsilon))$ by \cref{spin:prop:ising-work}.
The weight-preserving projection and spin recovery contract TV distance.
Adding the preprocessing and isolated-spin costs from
\cref{sec:computational-model} gives the stated bound on the original graph.
\end{proof}

\section{XY Sampling up to and at the Robust Threshold}\label{spin:sec:xy}

Consider the conserved integer-flow law \eqref{spin:eq:xy-flow} on a
simple graph $G$ of maximum degree at most $\Delta\ge3$.
We record edge multiplicities and transitions between consecutive
edges. Distinct polymers may share vertices, although they cannot
share physical edges. At each shared vertex, interval marks order
their auxiliary visit times; integrating these marks supplies the
factorial required by the flow weight.
The representation is exact before any temperature restriction.
In the robust-threshold regime, the exploration supplies spectral
and activity-tail bounds; a last-exit correction turns it into an
exact proposal procedure. We then recover spins and account for
the work of simulating the insertion--deletion chain.

\subsection{The marked polymer representation}\label{spin:app:xy-representation}
\begin{definition}[Marked polymers for integer flows]\label{spin:def:xy-polymer}
Let $\vec E$ contain both orientations of each physical edge.
A base type $P=(B,T)$ consists of a nonempty $B\subseteq\vec E$ and a
nonnegative integer matrix $T=(t_{ab})_{a,b\in B}$ such that:
\begin{enumerate}[label=(\roman*)]
\item $B$ uses at most one orientation of each physical edge, and
$t_{ab}=0$ unless $\operatorname{head}(a)=\operatorname{tail}(b)$;
\item the row and column sums at each $a\in B$ agree in a positive
integer $m_a$;
\item the undirected graph on $B$ with edges $ab$ whenever
$t_{ab}+t_{ba}>0$ is connected.
\end{enumerate}
Here $t_{ab}$ counts transitions from directed edge $a$ to directed
edge $b$. Equivalently, $P$ records the multiplicities and cyclic
transition counts of a closed Euler word, without specifying the
order of its individual traversals. Let $F_P\subseteq E$ be its
physical-edge support, $U_P=V(F_P)$ its vertex support, and
$n_v(P)=\sum_{a:\operatorname{tail}(a)=v}m_a$ its number of departures
from $v$. Write $|P|=\sum_a m_a$ for its total traversal length.

For $d\ge1$, define a finite measure $\Lambda_d$ by
\begin{equation}\label{spin:eq:xy-simplex-marks}
 \int f(s,t)\,d\Lambda_d(s,t)
 =\int_{0<x_1<\cdots<x_d<1}
       f(x_1,x_d)\,dx_1\cdots dx_d\qquad(f\ge0).
\end{equation}
Its total mass is $1/d!$. The marks $s,t$ are the first and last
of the $d$ ordered auxiliary visit times.
A marked polymer is $\gamma=(P,(s_v,t_v)_{v\in V(F_P)})$, where
$0<s_v\le t_v<1$. Its activity measure is
\begin{equation}\label{spin:eq:xy-activity}
 d\omega(\gamma)=
 \frac{\prod_{a\in B}(m_a!)^2}{\prod_{a,b\in B}t_{ab}!}
 \prod_{a\in B}r_{m_a}
 \prod_{v\in V(F_P)}d\Lambda_{n_v(P)}(s_v,t_v).
\end{equation}
Fix an order of the physical edges, and write $P\prec Q$ if the
least edge of $F_P$ precedes that of $F_Q$. Marked polymers sharing
a physical edge are incompatible. For edge-disjoint base types,
order them as $P\prec Q$. Then $\gamma=(P,(s_v,t_v))$ and
$\gamma'=(Q,(s'_v,t'_v))$ satisfy
\[
 \gamma\sim\gamma'\quad\Longleftrightarrow\quad
 t_v<s'_v\ \text{ for every }v\in U_P\cap U_Q.
\]
Thus all visits of the earlier polymer precede those of the later
one at each shared vertex.
\end{definition}

\begin{example}[Two flow polymers sharing a vertex]
\label{spin:ex:xy-marks}
To illustrate \cref{spin:def:xy-polymer}, consider the two directed
triangles below. The blue base type $P$ traverses its triangle twice,
whereas the orange base type $Q$ traverses its triangle once. Choose
the physical-edge order so that $P\prec Q$. They share the vertex
$v$, but no physical edge; whether their marked versions are
compatible depends on the marks at $v$.

\begin{center}
\begin{minipage}[t]{0.48\textwidth}
\centering
\textbf{(a) Edge directions and multiplicities}\par\smallskip
\begin{tikzpicture}[>=Stealth,scale=0.9,
 vertex/.style={circle,fill=black,inner sep=1.7pt},
 blueflow/.style={->,very thick,blue!65!black},
 orangeflow/.style={->,thick,dashed,orange!85!black}]
 \coordinate (v) at (0,0);
 \coordinate (u) at (-2,1.3);
 \coordinate (w) at (-2,-1.3);
 \coordinate (z) at (2,1.3);
 \coordinate (y) at (2,-1.3);
 \draw[blueflow] (v)--node[above,sloped] {$a_1\ (2)$}(u);
 \draw[blueflow] (u)--node[left] {$a_2\ (2)$}(w);
 \draw[blueflow] (w)--node[below,sloped] {$a_3\ (2)$}(v);
 \draw[orangeflow] (v)--node[above,sloped] {$b_1\ (1)$}(z);
 \draw[orangeflow] (z)--node[right] {$b_2\ (1)$}(y);
 \draw[orangeflow] (y)--node[below,sloped] {$b_3\ (1)$}(v);
 \node[blue!65!black] at (-1.3,0) {$P$};
 \node[orange!85!black] at (1.3,0) {$Q$};
 \foreach \p/\pos in {v/below,u/above,w/below,z/above,y/below}
  \node[vertex,label=\pos:{$\p$}] at (\p) {};
\end{tikzpicture}
\par\small Parentheses give the traversal multiplicity.
\end{minipage}\hfill
\begin{minipage}[t]{0.49\textwidth}
\centering
\textbf{(b) Auxiliary visit times at $v$}\par\smallskip
\begin{tikzpicture}[>=Stealth,x=4.5cm,y=1cm,
 visit/.style={circle,fill=blue!65!black,inner sep=2pt},
 qvisit/.style={rectangle,fill=orange!85!black,inner sep=2.3pt}]
 \draw[->,gray] (0,0)--(1.04,0);
 \node[below] at (0,0) {$0$};
 \node[below] at (1,0) {$1$};
 \draw[line width=3pt,blue!65!black] (0.2,0)--(0.48,0);
 \node[visit,label=above:{$x_1=s_v$}] at (0.2,0) {};
 \node[visit,label=below:{$x_2=t_v$}] at (0.48,0) {};
 \node[qvisit,label=above:{$y=s'_v=t'_v$}] at (0.8,0) {};
 \node[anchor=west,font=\small] at (0,-0.7)
 {Compatible: $x_1<x_2<y$.};
 \draw[->,gray] (0,-1.5)--(1.04,-1.5);
 \node[below] at (0,-1.5) {$0$};
 \node[below] at (1,-1.5) {$1$};
 \draw[line width=3pt,blue!65!black] (0.2,-1.5)--(0.8,-1.5);
 \node[visit,label=above:{$x_1=s_v$}] at (0.2,-1.5) {};
 \node[visit,label=above:{$x_2=t_v$}] at (0.8,-1.5) {};
 \node[qvisit,label=below:{$y=s'_v=t'_v$}] at (0.48,-1.5) {};
 \node[anchor=west,font=\small] at (0,-2.3)
 {Incompatible: $x_1<y<x_2$.};
\end{tikzpicture}
\end{minipage}
\end{center}

For $P$, the directed-edge set is $B=\{a_1,a_2,a_3\}$, and the
cyclic word $a_1a_2a_3a_1a_2a_3$ gives
\[
 T=\begin{pmatrix}0&2&0\\0&0&2\\2&0&0\end{pmatrix},
 \qquad m_{a_1}=m_{a_2}=m_{a_3}=2,
 \qquad |P|=6.
\]
Rows and columns are indexed by $a_1,a_2,a_3$. For instance,
$t_{a_1a_2}=2$ counts the two transitions from $a_1$ to $a_2$;
the cyclic transition from the last edge back to the first is included.
Every row and column sums to two, and the transition graph on $B$
is the connected cycle $a_1a_2a_3$.
The physical support is $F_P=\{uv,uw,wv\}$, its vertex support is
$U_P=\{u,v,w\}$, and $n_u(P)=n_v(P)=n_w(P)=2$.
For $Q$, each multiplicity and cyclic transition count is one,
$|Q|=3$, and $n_v(Q)=1$.

At the shared vertex, $P$ has two ordered auxiliary times
$0<x_1<x_2<1$, with mark $(s_v,t_v)=(x_1,x_2)$, whereas $Q$
has one time $y$, with mark $(s'_v,t'_v)=(y,y)$.
These are auxiliary coordinates, not the elapsed times of a walk;
the endpoint $t_v$ is also distinct from the integer transition
counts $t_{ab}$. Marks at the other vertices impose no constraints
between this pair. Since $P\prec Q$, compatibility is precisely
$t_v<s'_v$, or $x_2<y$. Merely sharing a vertex therefore does not
make the polymers incompatible, and disjoint intervals in the reverse
order would not suffice.

This example also explains why the mark measures are not normalized
probability measures. Before imposing compatibility, the product
mark measure at $v$ has total mass $1/(2!\,1!)$; after imposing it,
\[
 \int\one_{\{t_v<s'_v\}}\,
 d\Lambda_2(s_v,t_v)\,d\Lambda_1(s'_v,t'_v)
 =\int_{0<x_1<x_2<y<1}dx_1\,dx_2\,dy
 =\frac1{3!}.
\]
Thus ordering the marks supplies the factorial for all three visits
at the shared vertex, rather than separate factorials for the two
polymers. The same concatenation is used in the projection argument
below.
\end{example}

The total activity of all marks on a base type is
\begin{equation}\label{spin:eq:xy-base-weight}
 w_P=\frac{\prod_a(m_a!)^2\prod_a r_{m_a}}
             {\prod_{a,b}t_{ab}!\prod_v n_v(P)!}.
\end{equation}
To see the role of the marks, fix a vertex shared by base types
$P_1\prec\cdots\prec P_q$, and put $d=\sum_i n_v(P_i)$. Their
compatibility constraints concatenate the ordered visit times, so
\[
 \int\one_{\{t_1<s_2,\ldots,t_{q-1}<s_q\}}
       \prod_{i=1}^q d\Lambda_{n_v(P_i)}(s_i,t_i)=\frac1{d!}.
\]
This is the vertex factorial used in the projection identity below.
Project a compatible family to a flow by putting multiplicity $m_a$
in direction $a$ on its used edges and zero on unused edges, with
signs taken relative to the reference orientation.

\begin{proposition}[Exact flow projection]\label{spin:prop:xy-projection}
For every conserved flow $j$, the total product activity of compatible
marked families projecting to $j$ is $\prod_e r_{|j_e|}$. Consequently,
the polymer partition function equals $Z_{\mathbb Z}$, and the
projected normalized law is \eqref{spin:eq:xy-flow}.
\end{proposition}

\begin{proof}[Proof of \cref{spin:prop:xy-projection}]
\phantomsection\label{proof:spin:prop:xy-projection}
Fix a conserved flow $j$, orient its nonzero edges in the flow
direction, and put $m_a=|j_{e(a)}|$. At each vertex $v$, the incoming
and outgoing multiplicities both sum to $d_v$. Consider all transition
matrices $T$ with these row and column sums and with
$t_{ab}=0$ unless $\operatorname{head}(a)=\operatorname{tail}(b)$.
Their connected components are precisely the edge-disjoint base
types of families projecting to $j$.

At a vertex, order the components by their least physical edge and
write their visit counts as $n_1,\ldots,n_q$. By
\eqref{spin:eq:xy-simplex-marks}, each $\Lambda_{n_i}$ integrates
the first and last coordinates of an ordered simplex.
Compatibility places all visits of one component before those of
the next, so
\[
 \int\one_{\{t_1<s_2,\ldots,t_{q-1}<s_q\}}
       \prod_{i=1}^q d\Lambda_{n_i}(s_i,t_i)
 =\int_{0<x_1<\cdots<x_{d_v}<1}\prod_{i=1}^{d_v}dx_i
 =\frac1{d_v!}.
\]
Thus the contribution from a fixed $T$ is
\[
 \left(\prod_a r_{m_a}\right)
 \frac{\prod_a(m_a!)^2}{\prod_{a,b}t_{ab}!}
 \prod_v\frac1{d_v!}.
\]

For incoming multiplicities $p_1,\ldots,p_s$ and outgoing
multiplicities $q_1,\ldots,q_t$, both summing to $d$, the multinomial
identity gives
\[
 \sum_{\substack{t_{ij}\ge0\\\sum_jt_{ij}=p_i,\ \sum_it_{ij}=q_j}}
 \frac{\prod_i p_i!\prod_j q_j!}{\prod_{i,j}t_{ij}!}=d!.
\]
The constraints at different physical vertices use disjoint matrix
positions, so multiplying these identities gives
\[
 \sum_T\frac{\prod_a(m_a!)^2}{\prod_{a,b}t_{ab}!}
 =\prod_v d_v!.
\]
The preceding contribution therefore sums to $\prod_e r_{|j_e|}$.
Tonelli justifies the nonnegative sums and integrals. Finally,
$Z_{\mathbb Z}\le(\beta/I_0(\log\beta))^m<\infty$, by summing all edge values
without the conservation constraint. Normalization proves the claim.
\end{proof}

\subsection{Bessel-ratio exploration}\label{spin:app:xy-exploration}
From here through the activity-tail estimates, assume
$1<\beta\le\beta_{\rm XY}(\Delta)$, or equivalently
$(\Delta-1)r_1\le1$, with $\Delta\ge3$.
Once a physical edge is first traversed, its direction is fixed:
later traversals may use that direction but never its reverse.
Let $k_{vu}$ count traversals from $v$ to $u$, and let
$j_v=1+\sum_u k_{vu}$ be one plus the number of previous departures
from $v$. A fresh edge receives weight $r_1$; repeated use changes
the edge weight by a Bessel ratio.

\begin{enumerate}[label=\textbf{\arabic*.},leftmargin=*]
\item \textbf{Initialize.} Set the current vertex to $v=s$, all directed
traversal counts to $k_{uv}=0$, and all departure indices to $j_u=1$.
The stored path initially consists of $s$.
\item \textbf{Form the next row.} At the current vertex $v$, assign
to each neighbor $u$ the probability
\begin{equation}\label{spin:eq:xy-increment}
 p(v,u)=
 \begin{cases}
 r_1/j_v,&k_{vu}=k_{uv}=0,\\[2pt]
 \displaystyle\frac{(k_{vu}+1)^2}{j_v k_{vu}}
                   \frac{r_{k_{vu}+1}}{r_{k_{vu}}},&k_{vu}>0,\\[6pt]
 0,&k_{uv}>0.
 \end{cases}
\end{equation}
On the first step of the entire path, divide this row by
$\max\{1,\Delta r_1\}$.
\item \textbf{Move or stop.} Choose $u$ with probability $p(v,u)$,
or stop with probability $1-\sum_{u\sim v}p(v,u)$.
If moving, increment $j_v$ and $k_{vu}$, append $u$ to the path,
set $v=u$, and return to Step~2. If stopping, return the stored path
and its traversal counts.
\end{enumerate}

Under $(\Delta-1)r_1\le1$, the Bessel estimates in
\cref{spin:lem:xy-ratios} make every row substochastic and ensure
almost-sure termination.

\paragraph{Extracting a terminal closed word.}
For a nonempty closed directed-edge word $C=(a_1,\ldots,a_\ell)$,
put $a_{\ell+1}=a_1$ and define
\begin{equation}\label{spin:eq:xy-word-type}
 B(C)=\{a_1,\ldots,a_\ell\},\qquad
 t_{ab}(C)=|\{i\in[\ell]:(a_i,a_{i+1})=(a,b)\}|,\qquad
 P(C)=(B(C),T(C)).
\end{equation}
In particular, the transition from the last edge to the first is
included. Every closed word produced by the exploration has a base
type $P(C)$ as in \cref{spin:def:xy-polymer}.

Write the stopped path as $v_0,\ldots,v_L$ and set
\[
 i_* =\min\{0\le i\le L:v_i=v_L\}.
\]
If $i_*<L$, extract the terminal word
$C_{\rm end}=((v_i,v_{i+1}))_{i=i_*}^{L-1}$; if $i_*=L$, there is
no nonempty terminal word. This rule uses the first visit to the
terminal vertex in the entire path. It will define the completion
events for the first-hit analysis. The exact proposal generator
instead retains only paths closing at their starting vertex and
then corrects their output probabilities (\cref{spin:lem:xy-oracle}).

The completion estimate below uses the constant
\begin{equation}\label{spin:eq:xy-kappa}
 \kappa=\frac{2\max\{1,\Delta r_1\}}{1-\log\beta+r_1}.
\end{equation}

\begin{lemma}[Bessel bounds for the exploration]\label{spin:lem:xy-ratios}
The normalized Bessel weights satisfy
\[
 (k+1)\frac{r_{k+1}}{r_k}\le \log\beta/2\quad(k\ge0),
 \qquad \log\beta\le\frac{2r_1}{1-r_1^2}.
\]
If $\Delta\ge3$ and $(\Delta-1)r_1\le1$, the continuation mass at the
$j$th departure after the initial step is at most
\[
 \frac{(\Delta-1)r_1+(j-1)(\log\beta-r_1)}j
 \le\frac{1+(j-1)(\log\beta-r_1)}j.
\]
The exploration in \cref{spin:sec:xy} is well defined and terminates
almost surely. Its constant in \eqref{spin:eq:xy-kappa} satisfies
\[
 \kappa\le K_\Delta:=
 \frac{2\Delta^2(\Delta-2)}{\Delta(\Delta-2)^2-2}.
\]
The bound $K_\Delta$ decreases with $\Delta$ and is strictly below
$e$ for $\Delta\ge8$.
\end{lemma}

The Bessel inputs are the series and recurrence formulas
\cite[Sections~10.25 and~10.29]{DLMFSpins} and the Tur\'an inequality
\cite{BariczPonnusamy13}.

\begin{proof}[Proof of \cref{spin:lem:xy-ratios}]
\phantomsection\label{proof:spin:lem:xy-ratios}
The Bessel series gives $(k+1)I_{k+1}(\log\beta)\le(\log\beta/2)I_k(\log\beta)$ term by
term. Tur\'an's inequality gives $r_2\le r_1^2$, and the recurrence
$1-r_2=2r_1/\log\beta$ then gives $\log\beta\le2r_1/(1-r_1^2)$.
In particular, $r_1\le1/2$ implies $\log\beta-r_1\le5/6$.

At a departure after the initial step, the incoming edge is forbidden.
At most $j-1$ outgoing edges have been used, and a used edge of count
$k$ contributes
\[
 \frac{(k+1)^2}{k}\frac{r_{k+1}}{r_k}\le \log\beta
\]
before division by $j$. The other available edges contribute $r_1$.
This gives the stated continuation bound. At each repeat departure
the stopping probability is at least
$(1-\log\beta+r_1)/2\ge1/12$. A path has at most $n$ first departures, so its
natural length has tail at most $(11/12)^{(L-n)_+}$ and is finite
almost surely.

Put $s=1/(\Delta-1)$. Since $r_1\le s$, the Bessel bound gives
$\log\beta-r_1\le s(1+s^2)/(1-s^2)$, while
$\max\{1,\Delta r_1\}\le1+s$. Thus \eqref{spin:eq:xy-kappa} gives
\[
 \kappa\le\frac{2(1+s)}{1-s(1+s^2)/(1-s^2)}
 =\frac{2\Delta^2(\Delta-2)}{\Delta(\Delta-2)^2-2}=K_\Delta.
\]
The numerator of the first expression increases with $s$, and its
positive denominator decreases with $s$ for $0<s\le1/2$.
Hence $K_\Delta$ decreases with $\Delta$, and for $\Delta\ge8$,
$K_\Delta\le K_8=384/143<e$.
\end{proof}

\paragraph{Counting completions.}
The next count sums the probabilities of all closed words with a
given base type. It supplies the completion bound for first hits and
will also identify the correction needed by the proposal algorithm.

\begin{lemma}[Euler-word weights]\label{spin:lem:xy-euler}
For a base type $P=(B,T)$, let $\tau(T)$ be the common number of
spanning in-trees rooted at each state of the Eulerian transition
multigraph, with parallel transitions distinguished. For $a\in B$,
the number $N_a(T)$ of rooted words starting with $a$ and having
cyclic transition counts $T$ is
\[
 N_a(T)=\frac{m_a\tau(T)\prod_{b\in B}(m_b-1)!}
                    {\prod_{b,c\in B}t_{bc}!}.
\]
For $N_v(P)=\sum_{a:\operatorname{tail}(a)=v}N_a(T)$,
\[
 N_v(P)\frac{\prod_a m_a m_a!r_{m_a}}{\prod_u n_u(P)!}
 =n_v(P)\tau(T)w_P\ge n_v(P)w_P.
\]
\end{lemma}

\begin{proof}[Proof of \cref{spin:lem:xy-euler}]
\phantomsection\label{proof:spin:lem:xy-euler}
Use the rooted Euler-tour formula \cite{AardenneBruijn51} on the
transition multigraph. For each labelled in-tree toward $a$, put its
edge last in the outgoing order at every nonroot state and order the
other transitions freely. Following these orders gives every labelled
rooted tour once. There are
$m_a\tau(T)\prod_b(m_b-1)!$ such tours. Each rooted word has
$\prod_{b,c}t_{bc}!$ transition labellings, giving the formula for
$N_a(T)$.

Balance and connectivity make the directed transition graph strongly
connected. The directed matrix-tree theorem gives equal rooted in-tree
counts $\tau(T)\ge1$. Substituting into $N_v(P)$ proves the second formula.

\end{proof}

\subsection{First-hit bound and conflict spectrum}

The preceding exploration supplies the following conflict bound.

\begin{proposition}[The marked-polymer first-hit bound]\label{spin:prop:xy-hit}
If $\Delta\ge3$ and $(\Delta-1)r_1\le1$, the marked polymer model
admits a witness, constant on base types with the same vertex support,
satisfying
\[
 AWh\le\kappa h,\qquad 1\le h_P\le n,
 \qquad \int h\,d\omega\le\kappa n.
\]
In particular, $\lambda_{\max}(AW)\le\kappa$.
\end{proposition}
\begin{proof}
For a vertex set $U$, let $\tau_U=\inf\{0\le i\le L:v_i\in U\}$
and $H_U=\{\tau_U\le L\}$. For a base type $P=(B,T)$, define
\begin{equation}\label{spin:eq:xy-completion-event}
 E_P=\{i_*<L,\ \tau_{U_P}=i_*,\ P(C_{\rm end})=P\}.
\end{equation}
Thus completion starts at the first visit to $U_P$, traverses a
closed word of type $P$, and stops upon its final return. It completes
a base type; no interval marks are drawn in this experiment.

Fix any history $\mathcal H$ up to the first visit to $U_P$, and let
$v=v_{\tau_{U_P}}$. Before the next move-or-stop choice,
\[
 k_{xy}=k_{yx}=0\quad(xy\in F_P),\qquad
 j_u=1\quad(u\in U_P).
\]
If the path entered from outside $U_P$, its entering physical edge
is not in $F_P$. The prohibition on reversing a used edge therefore
excludes no edge required by $P$.

Let $\mathcal W_v(P)$ be all closed directed-edge words $C$ with
$P(C)=P$ and $\operatorname{tail}(a_1)=v$. Each has length $|P|$.
The Euler-word count in \cref{spin:lem:xy-euler} gives
\[
 \sum_{C\in\mathcal W_v(P)}
 \Prb_s(\text{next }|P|\text{ traversals form }C\mid\mathcal H)
 \ge\frac{n_v(P)w_P}{\max\{1,\Delta r_1\}}.
\]
After this word, the path is back at $v$ with departure index
$n_v(P)+1$. By \cref{spin:lem:xy-ratios}, it stops there with
probability at least $n_v(P)(1-\log\beta+r_1)/(n_v(P)+1)$. Hence
\[
 \Prb_s(E_P\mid\mathcal H)
 \ge\frac{1-\log\beta+r_1}{\max\{1,\Delta r_1\}}
       \frac{n_v(P)^2}{n_v(P)+1}w_P
 \ge\frac{w_P}{\kappa}.
\]
Averaging over first-hit histories gives
$w_P\Prb_s(H_{U_P})\le\kappa\Prb_s(E_P)$.

The stopped path determines $i_*$, $C_{\rm end}$, and its unique
base type. Thus the events $E_P$ are mutually exclusive. To apply
\cref{pol:lem:first-hit} to distinct vertex supports, set
\[
 w_U=\sum_{P:U_P=U}w_P,\qquad
 E_U=\bigsqcup_{P:U_P=U}E_P,\qquad
 h_\gamma=h_P=\sum_{s\in V}\Prb_s(H_{U_P}).
\]
Summing the preceding inequality over $P$ with support $U$ gives
$w_U\Prb_s(H_U)\le\kappa\Prb_s(E_U)$. Completion visits every
vertex of $U$, so $E_U\subseteq H_{U'}$ whenever $U\cap U'\ne\varnothing$.
Every $s\in U$ is a certain-hitting start, and every marked-polymer
conflict implies vertex intersection. The first-hit lemma with
$\mathcal S=V$ now gives the claimed bounds.
\end{proof}

For $\Delta\ge8$ and $(\Delta-1)r_1\le1$, the Bessel estimates in
\cref{spin:lem:xy-ratios} give $\kappa\le384/143<e$. This verifies
the upper-spectrum condition of \cref{pol:lem:sampling}.

\subsection{Exact proposals and activity tails}
The spectral certificate used completion after the first hit of a
support. For proposals we keep the entire closed path rooted at its
starting vertex. We first remove the multiplicity of words representing
the same base type, then draw its interval marks.

For the proposal correction, regard the closed words in
\eqref{spin:eq:xy-word-type} modulo cyclic rotation, retaining their
orientation. The transition matrix $T(C)$ is unchanged by rotation.
Let $s(C)$ be the number of rotations fixing $C$; a word of length
$L$ therefore has $L/s(C)$ distinct rooted representatives. Define
its auxiliary activity
\begin{equation}\label{spin:eq:word-activity}
 \widehat w_C=\frac1{s(C)}
 \frac{\prod_a m_a m_a!r_{m_a}}{\prod_v n_v!}.
\end{equation}

The rooted count in \cref{spin:lem:xy-euler} also gives the identity
needed for the proposal correction. A cyclic word has $m_a/s(C)$
distinct roots at state $a$, so
\begin{equation}\label{spin:eq:word-fiber}
 \sum_{C:T(C)=T}\widehat w_C=\tau(T)w_P.
\end{equation}
All words in this sum have the same physical support as $P$.

\begin{lemma}[Last-exit correction]\label{spin:lem:last-exit}
Conditional on a transition matrix $T$, suppose a cyclic word $C$
has probability proportional to $1/s(C)$. There is a test using
$O(L)$ operations on a word of length $L$, with acceptance probability
$1/\tau(T)$.
\end{lemma}

\begin{proof}[Proof of \cref{spin:lem:last-exit}]
\phantomsection\label{proof:spin:lem:last-exit}
Choose a state $a_0$ deterministically from $T$, and choose a spanning
in-tree $R$ toward it by reverse breadth-first search. Select uniformly
one occurrence of $a_0$ in the proposed cyclic word and open the word
there. Each distinct rooted word has $s(C)$ preimages, cancelling its
$1/s(C)$ weight; the rooted word is therefore uniform.

Give occurrences of each transition $a\to b$ uniform distinct labels
for the counting argument. Every rooted word has the same number of
labellings. The last-exit bijection in the proof of \cref{spin:lem:xy-euler} shows that
its last-exit tree is uniform among the $\tau(T)$ labelled in-trees.
Accept when it equals $R$, with label one on each selected transition.

The test can be performed without storing these labels: check that
the last departure from each $a\ne a_0$ goes to its parent $R(a)$,
then independently accept with probability $1/t_{a,R(a)}$ at each
such state. These probabilities integrate out the labels exactly.
The tree search and word scan take $O(L)$ operations.
\end{proof}

\begin{lemma}[Exact marked proposals]\label{spin:lem:xy-oracle}
Suppose $\Delta\ge3$ and $(\Delta-1)r_1\le1$. There is a trial
based on the XY exploration whose candidate law is
\[
 \Prb(\text{output in }d\gamma)=\frac{d\omega(\gamma)}{\kappa n}.
\]
Successful outputs have the representation in \cref{spin:def:xy-polymer}.
The expected exploration length is $O(\log n)$, including unsuccessful
trials. For $N\ge\kappa n$, running the trial with probability
$\kappa n/(4N)$ implements the insertion proposal of $P$.
\end{lemma}

\begin{proof}[Proof of \cref{spin:lem:xy-oracle}]
\phantomsection\label{proof:spin:lem:xy-oracle}
Start the exploration at a uniform physical vertex. For a fixed
rooted closed word of length $L$, the edge increments and departure
denominators telescope to give continuation probability
\[
 \frac1{\max\{1,\Delta r_1\}}
 \frac{\prod_a m_a m_a!r_{m_a}}{\prod_v n_v!}
 =\frac{s(C)\widehat w_C}{\max\{1,\Delta r_1\}}.
\]
Impose an independent horizon with survival probability $1/L$ at
length $L$. At each endpoint, test natural termination first. If it
does not occur, abort with probability $1/(L+1)$ before taking the
next step. Retain a candidate only when natural termination occurs
at the starting vertex after a nonempty closed word.
If the termination probability there is $d$, retain it further with
probability $(1-\log\beta+r_1)/(2d)\le1$, using \cref{spin:lem:xy-ratios}.

Each rooted word now has probability
$s(C)\widehat w_C/(\kappa n L)$. There are $L/s(C)$ distinct roots,
so the cyclic word has probability $\widehat w_C/(\kappa n)$.
Apply \cref{spin:lem:last-exit}. By \eqref{spin:eq:word-fiber}, the
resulting base type has probability $w_P/(\kappa n)$. Draw its marks
by taking the minimum and maximum of $n_v(P)$ independent uniform
points at each visited vertex. This is exactly the normalized mark
law $\prod_{v\in U_P}n_v(P)!\,d\Lambda_{n_v(P)}$ conditional on $P$.
Since $\Lambda_d$ has mass $1/d!$, this gives precisely the activity
measure in \eqref{spin:eq:xy-activity}, divided by $\kappa n$.

The natural length tail and the horizon give expected exploration length
at most
\[
 \sum_{L\ge1}\frac{(11/12)^{(L-n)_+}}L=O(\log n).
\]
All corrections and mark generation take work linear in the returned
word's length. Running the trial with probability $\kappa n/(4N)$
implements the insertion rule of $P$.
\end{proof}

\paragraph{Activity tails.}
The proposal also controls total traversal length and conflicting
activity. Both estimates include the full Bessel multiplicities.

\begin{lemma}[Length and conflicting-activity tails]\label{spin:lem:xy-tail}
If $\Delta\ge8$ and $(\Delta-1)r_1\le1$, then
\[
 \int |\gamma|\,d\omega\le\kappa n,
 \qquad
 \int |\gamma|(3/2)^{D(\gamma)/\kappa}
 \,d\omega(\gamma)\le2\kappa n.
\]
\end{lemma}

\begin{proof}[Proof of \cref{spin:lem:xy-tail}]
\phantomsection\label{proof:spin:lem:xy-tail}
Omit the independent horizon from the preceding proposal, retaining
both corrections. The same calculation gives the length-biased law
\begin{equation}\label{spin:eq:length-proposal}
 \Prb(\text{output in }d\gamma)
 =\frac{\sum_a m_a}{\kappa n}\,d\omega(\gamma).
\end{equation}
Its total mass is at most one, proving the first bound.

For the exponential bound, use the auxiliary word activities in
\eqref{spin:eq:word-activity}. For a cyclic word $C$, write $U_C$ for its vertex support and use
the first-hit time $\tau_{U_C}$ and terminal word $C_{\rm end}$ from
\cref{spin:sec:xy}. Define
\[
 H_C=\{\tau_{U_C}\le L\},\qquad
 E_C=\{i_*<L,\ \tau_{U_C}=i_*,\
                [C_{\rm end}]_{\rm cyc}=C\},
\]
where $[\,\cdot\,]_{\rm cyc}$ denotes the cyclic-rotation class.
The first-hit argument, before the last-exit correction, gives
\[
 \Prb(E_C\mid\text{first-hit history})\ge\widehat w_C/\kappa.
\]
Indeed, there are $n_v(C)/s(C)$ distinct roots at the first-hit vertex
$v$; their continuation probabilities telescope as above, and the
final stop has probability at least
$n_v(C)(1-\log\beta+r_1)/(n_v(C)+1)$. The completed suffix determines $C$, so
these events are disjoint, as in \cref{spin:prop:xy-hit}.

Let $Y$ be the total auxiliary activity of words whose vertex
supports the natural exploration visits. Starting fresh at $v$ gives
$\sum_{C:v\in V(C)}\widehat w_C\le\kappa$.
After any history, words not yet hit have fresh support at their
future first hit. The same disjoint-completion argument bounds the
conditional expected additional activity by $\kappa$.

Let $x_j$ be the activity first discovered at the $j$th newly visited
vertex, padded with zeros to length $n$. Then $0\le x_j\le\kappa$,
$Y=\sum_jx_j$, and the conditional expected remaining sum is at most
$\kappa$. Put $q_j=(3/2)^{x_j/\kappa}-1\le x_j/(2\kappa)$.
Repeated conditioning on the penultimate discovery gives
\[
 \E\sum_{j_1<\cdots<j_k}q_{j_1}\cdots q_{j_k}\le2^{-k},
 \qquad
 \E(3/2)^{Y/\kappa}\le\sum_{k\ge0}2^{-k}=2.
\]
By \eqref{spin:eq:word-fiber}, total word activity dominates total
marked activity on each support. Every marked conflict implies vertex
intersection. On an output $\gamma$ of the length-biased proposal,
the visited support is precisely its support, so $D(\gamma)\le Y$.
Using \eqref{spin:eq:length-proposal}, we obtain
\[
 \frac1{\kappa n}\int\Bigl(\sum_a m_a\Bigr)
       (3/2)^{D(\gamma)/\kappa}\,d\omega
 \le\E[\one_{\{\mathrm{output}\}}(3/2)^{Y/\kappa}]\le2.
\]
Finite word truncations and monotone convergence justify the
countable sums throughout.
\end{proof}

\subsection{Spin recovery and work}

We pass from integer flows through directed currents and colored
loops to two conditional Ising models.

\begin{lemma}[Recovery of XY spins]\label{spin:lem:xy-recovery}
There is a randomized kernel sending the integer-flow law
\eqref{spin:eq:xy-flow} to the XY law. It contracts TV distance. In the range of \cref{spin:thm:xy},
its conditional expected work on input $j$ is
$O(m+\sum_e|j_e|)$.
\end{lemma}

\begin{proof}[Proof of \cref{spin:lem:xy-recovery}]
\phantomsection\label{proof:spin:lem:xy-recovery}
On each edge independently draw
\[
 \Prb(k_e=k\mid j_e)=
 \frac{(\log\beta/2)^{2k+|j_e|}}{k!(k+|j_e|)!\,I_{|j_e|}(\log\beta)},\qquad k\ge0,
\]
and set $N_e^+=k_e+\max\{j_e,0\}$ and
$N_e^-=k_e+\max\{-j_e,0\}$. The Bessel series cancels the flow
weights, giving the balanced directed-current law with weight
$\prod_e (\log\beta/2)^{N_e^++N_e^-}/(N_e^+!N_e^-!)$
\cite[Equations~(13)--(14)]{LiuZhangTanDeng23}.

Label the links on each edge and assign directions uniformly subject
to these counts. At each vertex choose a uniform bijection from
incoming to outgoing links. Forget directions, and color each resulting
loop independently with color 1 or 2. If the degrees of the two
colors at $v$ are $2d_v^1,2d_v^2$, independently draw
\[
 u_v\sim\operatorname{Beta}(d_v^1+1/2,d_v^2+1/2).
\]
Conditional on $u$, the two colored multiplicities have the laws of
independent sourceless currents for two zero-field Ising models with
edge activities
\[
 \beta^1_{uv}=\beta^{2\sqrt{u_uu_v}},\qquad
 \beta^2_{uv}=\beta^{2\sqrt{(1-u_u)(1-u_v)}}.
\]

In the random-wire representation \cite{BenassiUeltschi20}, summing
orientations of labelled links with multiplicities $m_e$ and a fixed
pairing gives
\[
 2^{\#\mathrm{loops}}\prod_e\frac{(\log\beta/2)^{m_e}}{m_e!}
                       \prod_v\frac1{(d_v^1+d_v^2)!}.
\]
Uniform loop coloring cancels the power of two. Summing same-color
pairings gives $(2d_v^1-1)!!(2d_v^2-1)!!$ at each vertex, and assigning
link colors gives the edge multinomial coefficients. The beta identity
\[
 \int_0^1u^{d_1}(1-u)^{d_2}\frac{du}{\pi\sqrt{u(1-u)}}
 =\frac{(2d_1-1)!!(2d_2-1)!!}{2^{d_1+d_2}(d_1+d_2)!}
\]
identifies this with the Cartesian expansion of the XY law: the
vertex factors $2^{d_1+d_2}$ supply a factor of two per link.
This gives the beta posterior and the conditional current laws.

For each color, keep every edge of positive current and independently
add each zero-current edge with probability $1-(\beta^c_{uv})^{-1/2}$.
The current-plus-Bernoulli coupling \cite{LupuWerner16} gives the
corresponding FK-Ising law. Color its clusters by independent fair
signs $\sigma^c$ and return the angles of
\[
 (\sigma_v^1\sqrt{u_v},\ \sigma_v^2\sqrt{1-u_v}),\qquad v\in V.
\]
Edwards--Sokal \cite{ES88} and the conditional current laws give the XY law.

To draw each cancelling count, propose $k\sim\operatorname{Poi}(\log^2\beta/4)$ and accept with
probability $|j_e|!/(|j_e|+k)!$. Acceptance
is at least $e^{-\log^2\beta/4}$; the factorial ratio is a product of $k$ factors.
The bounded value of $\log\beta$ makes the expected cost constant per edge.
The ratio of successive probabilities also gives
$\E[k_e\mid j_e]\le \log^2\beta/4$. Link pairing, loop coloring, beta sampling
and two component searches therefore take conditional expected work
$O(m+\sum_e|j_e|)$.
\end{proof}

\begin{proposition}[XY work]\label{spin:prop:xy-work}
Under the assumptions of \cref{spin:thm:xy}, and after the preprocessing
in \cref{sec:computational-model}, take $N=m+6n$ in
\cref{pol:lem:sampling}. The marked-polymer
chain from $\varnothing$ can be simulated for $t$
steps in
\[
 \wtO\!\left(m+\frac nN t\right)
\]
expected work. Spin recovery adds $O(m)$ expected work. The second
mixing horizon
in \cref{pol:lem:sampling} gives \cref{spin:thm:xy}.
\end{proposition}

\begin{proof}[Proof of \cref{spin:prop:xy-work}]
\phantomsection\label{proof:spin:prop:xy-work}
At the current vertex $v$, store the number $F$ of unused incident
edges, the list of used outgoing edges, and their total increment
weight. On departure $j_v$, the fresh branch has mass $F r_1/j_v$; conditional
on taking it, sample uniform incident edges until an unused one is
found. Its unconditional expected number of draws is
$(F r_1/j_v)\deg(v)/F\le\Delta r_1=O(1)$.
For the used branch, sample a uniform used outgoing edge and accept
with probability
$((k+1)^2/k)(r_{k+1}/r_k)/\log\beta$. Its unconditional expected number of
draws is at most $\log\beta$ times the number of used outgoing edges divided
by $j_v$, hence $O(1)$. Each traversal changes only the local data at
its endpoints. Lists of touched entries avoid repeated initialization.

The transition counts, last-exit test and endpoint marks use
$O(L\log n)$ work on a word of length $L$; dictionaries are indexed
by physical edges and ordered edge pairs. For compatibility, first
test physical-edge owners. At each visited vertex maintain an interval
tree keyed by the least physical edge of each present polymer.
A candidate interval
is compatible exactly when it lies after its predecessor's last mark
and before its successor's first mark. Tests and updates take
$O(L\log n)$ work; a later deletion is charged to the insertion.
Thus an attempted proposal, including its possible update and later
deletion, costs $O(\log^2 n)$ in expectation by the harmonic-horizon
bound in \cref{spin:lem:xy-oracle}.

In a state with $k$ present polymers, the next step that attempts a
proposal or deletes a present polymer occurs after a geometric wait
with parameter $(\kappa n+k)/(4N)$. At that step choose these two
actions in ratio $\kappa n:k$, deleting uniformly in the latter case.
Stop if the next step exceeds $t$. The expected number of attempted proposals
is $t\kappa n/(4N)$, and deletions are at most accepted insertions.
The resulting work is
$\wtO(m+(n/N)t)$.

For recovery, let $S_t$ be the total traversal length of the family
after $t$ steps. Each present polymer is deleted with probability
$1/(4N)$, and accepted insertions contribute at most the proposal
first moment. Therefore
\[
 \E S_{t+1}\le\left(1-\frac1{4N}\right)\E S_t
       +\frac1{4N}\int\Bigl(\sum_a m_a\Bigr)\,d\omega.
\]
Starting from zero, \cref{spin:lem:xy-tail} gives
$\E S_t\le\kappa n$ for every $t$. Since compatible polymers
have disjoint physical edges, $S_t=\sum_e|j_e|$ for the projected
flow. The conditional work bound in \cref{spin:lem:xy-recovery}
therefore averages to $O(m)$.
\end{proof}

\begin{proof}[Proof of \cref{spin:thm:xy}]
\phantomsection\label{proof:spin:thm:xy}
\Cref{spin:prop:xy-hit,spin:lem:xy-ratios} give
$\lambda_{\max}(AW)\le384/143<e$. Since
$\log(3/2)/\kappa>1/8$, the length-weighted tail in
\cref{spin:lem:xy-tail} implies \eqref{pol:eq:activity-tail} with
$N=m+6n$ and $\eta=1/8$. Every compatible family has at most $m$
members, so the global bounds also hold. Apply the upper-spectrum
case of \cref{pol:lem:sampling} and then \cref{spin:prop:xy-work}.
As $m\le n^2$, $\log(N/\varepsilon)=O(\log(n/\varepsilon))$;
the stated work follows. Exact flow projection and
\cref{spin:lem:xy-recovery} preserve the TV guarantee.
The preprocessing and independent isolated spins add $O(n+m)$ work
on the original input, which is absorbed by the stated bound.
\end{proof}

\section{Potts Sampling below the Order--Disorder Scale}\label{spin:sec:potts}
Put $p=\beta-1$, $r=p/(q+p)$, and
$p_{\od}(q,\Delta)=\beta_{\od}(q,\Delta)-1$.
The case $p=0$ is independent sampling, so assume $p>0$.
We use the $\mathbb Z_q$-flow representation from
\cref{spin:sec:flows}:
\[
 \pi_q(\xi)\propto\one_{\{\operatorname{div}\xi=0\}}
 r^{|\operatorname{supp}\xi|}.
\]
After summing over the internal flows of small connected regions,
we sample the exterior flow through its labelled component polymers.
Section~\ref{sec:holant-model} gives their first-hit criterion, and
Section~\ref{sec:regional-implementation} constructs the regions and
verifies the criterion for $p\le p_{\od}(q,\Delta)/10$.
The random-cluster coupling bounds and implements the regional
signature factors; it also recovers Potts spins from the sampled flow.
The final subsection accounts for these operations and the total work.

\subsection{Flow polymers and the first-hit criterion}\label{sec:holant-model}
\label{sec:holant-criterion}\label{sec:nonbinary-proof}\label{sec:holant}
Summing internal flows replaces vertex conservation constraints by
functions of the boundary labels. We formulate the component model and
its exploration criterion for these conserved signatures.

Let $H$ be a finite undirected loopless multigraph, and let $q\ge2$
be an integer. Each physical edge has two directed copies $a,\bar a$;
parallel edges retain distinct identities. Write $\vec E(H)$ for all
these copies and $E_v$ for those with tail $v$. A configuration
$\xi:\vec E(H)\to\mathbb Z_q$ is antisymmetric if
$\xi(\bar a)=-\xi(a)$. It is a flow if
$\sum_{a\in E_v}\xi(a)=0$ at every $v$.
For $q=2$, the two labels on an edge agree and the nonzero edges have
even degree at every vertex, giving exactly the even subgraphs.

Give each vertex a signature
$f_v:\mathbb Z_q^{E_v}\to\mathbb R_{\ge0}$ with $f_v(\mathbf0)>0$
and $f_v(x)=0$ whenever $\sum_{a\in E_v}x_a\ne0$.
Writing $\xi_v=(\xi(a))_{a\in E_v}$, define the weight
$w_f(\xi)=\prod_v f_v(\xi_v)$ on antisymmetric configurations,
the partition function $Z(H,f)=\sum_{\xi\text{ antisymmetric}}w_f(\xi)$,
and the law $\mu_{H,f}(\xi)=w_f(\xi)/Z(H,f)$.
Thus antisymmetry belongs to the edge labels, while conservation is a
support condition on the signatures. No symmetry, invariance under
$x\mapsto-x$, or factorization of the signatures is assumed.
The unweighted even-subgraph model at $q=2$ has even-parity signatures. Product edge weights can be split
between the endpoint signatures.

For calculations, fix one reference direction per physical edge and
identify $\xi$ with its labels in those directions. The label seen at
$v$ is then $\epsilon_{v,e}\xi_e$, where $\epsilon_{v,e}=1$ for an
outgoing reference edge and $-1$ for an incoming one.
Reversing a reference direction negates that coordinate and leaves
the outward labels unchanged. Write $\operatorname{supp}\xi$ for the
set of physical edges with nonzero label, counted once per edge.

\paragraph{Component polymers and contraction.}
A polymer $C=(F_C,\xi^C)$ is a positive-weight labelled connected
component of the nonzero support. Extend its labels by zero outside
$F_C$, and define
\begin{equation}\label{eq:holant-polymers}
 w_C=\prod_{v\in V(F_C)}\frac{f_v(\xi^C_v)}{f_v(0)},
 \qquad A_{CD}=\one\{V(F_C)\cap V(F_D)\ne\varnothing\}.
\end{equation}
Two polymers are incompatible when their vertex sets intersect,
so $A$ includes the diagonal, as in the abstract polymer convention
of \cref{pol:def:model}. The connected components give the exact law
$\mu_{\rm pol}(X)\propto\prod_{C\in X}w_C$ on vertex-disjoint
families. Every support is bridgeless: summing conservation on one
side of a bridge cancels the internal labels and forces the bridge
label to be zero. Write $W=\Diag(w)$.

The signature obtained by summing over the internal labels of a
connected region $R$ is
$F_R(\xi)=\sum_\eta\prod_{v\in R}f_v((\eta,\xi)_v)$,
where $\xi$ fixes the boundary labels and $\eta$ ranges over internal
antisymmetric labels. Summing conservation over $R$ cancels every
internal edge, so $F_R$ is again supported on boundary labels summing
to zero. Also $F_R(0)>0$. Hence contracting regions preserves this
class of Holant models, with parallel boundary edges kept distinct.
This closure property imposes no bound on $F_R(\xi)/F_R(0)$;
such bounds belong to the application.

For later use, define the divergence vector by
$(\operatorname{div}\xi)(v)=\sum_{a\in E_v}\xi(a)$ and let
$\Omega_b(H)$ be the antisymmetric configurations with divergence
$b\in\mathbb Z_q^{V(H)}$. The flow space is $\Omega_0(H)$.
Weighted sums over $\Omega_b(H)$ will be introduced with their
specific weights in \cref{sec:flow-coupling}.

\paragraph{The first-hit criterion.}

Return to a general Holant model with flow conservation on $H$, now with $q\ge3$.
For nonempty $S\subseteq\delta_H(v)$, let
$m_v(S)=\max_{\operatorname{supp}\zeta=S}f_v(\zeta)/f_v(0)$,
with value zero on unsupported inputs.

\begin{theorem}[Holant criterion for $q\ge3$]\label{thm:holant-criterion}
Let $q\ge3$ and put $Q=q-1$. Suppose there are $s\in[1,Q]$ and
positive incidence parameters $x_{v,e}$ satisfying
\begin{equation}\label{eq:holant-local-condition}
 \frac{s}{Q}
 \left[\prod_{e\ni v}\left(1+\frac Qs x_{v,e}\right)-1\right]\le1
 \qquad(v\in V(H)),
\end{equation}
and, for every nonempty connected bridgeless support $F$,
\begin{equation}\label{eq:holant-support-condition}
 Qs^{|F|-|V(F)|}\prod_{v\in V(F)}m_v(\delta_F(v))
 \le\prod_{e=uv\in F}\sqrt{x_{u,e}x_{v,e}}.
\end{equation}
Then there is a positive vector $h$ satisfying
\[
 AWh\le h/2,\qquad 1\le h_C\le |V(H)|,\qquad
 \sum_C w_C h_C\le |V(H)|/2.
\]
In particular, $\lambda_{\max}(AW)\le1/2$.
Equality is allowed in both certificate conditions.
\end{theorem}

The first condition allocates local probability mass. The second pays
for all conserved labellings of a support using products of those local
parameters. Neither condition requires enumerating the polymer space.
The proof uses independent outgoing-edge experiments in a depth-first
exploration. Completing a strong orientation after the first hit
forces its support to be the unique terminal component, allowing the
same first-hit lemma as in the Boolean case.

\begin{proof}[Proof of Theorem~\ref{thm:holant-criterion}]
\emph{1. Outgoing sets and strong orientations.}
Put $\kappa=Q/s$.
Independently at each $v$, select a nonempty outgoing set $O$ with
probability $\kappa^{|O|-1}\prod_{e\in O}x_{v,e}$; the remaining
probability is failure. The local condition is exactly its normalization.
For each bridgeless connected support $F$, fix a strong orientation
$F^+$ and its reversal $F^-$. Strong connectivity will ensure that
an exploration entering $F$ discovers every vertex of $F$.
Such a pair is obtained by orienting a depth-first tree away from the
root and every nontree edge towards its ancestor: each tree cut has
a nontree edge across it, so every vertex can also reach the root.
Parallel edges retain distinct identities.

\emph{2. Paying for all labels of a support.}
Let $B_F^\pm$ prescribe that, at each vertex of $F$, the selected
outgoing set is exactly the one in $F^\pm$. In particular, no selected
edge leaves $F$. Write $d_F^\pm=\Prb(B_F^\pm)$.
With $k=|V(F)|$ and $\ell=|F|$, multiplying the independent row
probabilities gives
\begin{equation}\label{eq:holant-event-budget}
 g_F:=\sqrt{d_F^+d_F^-}
 =\kappa^{\ell-k}\prod_{e=uv\in F}\sqrt{x_{u,e}x_{v,e}},
 \qquad W_F:=\sum_{C:F_C=F}w_C\le g_F.
\end{equation}
The geometric mean uses both endpoint parameters of each edge,
since reversing the orientation exchanges its tail and head.
For the inequality, first choose nonzero labels on the
$\ell-k+1$ chords of a spanning tree. Conservation then fixes the
tree labels by successive leaf elimination. Some of those labels may
be zero or forbidden, so this gives an upper bound of
$Q^{\ell-k+1}$ labellings, also for composite $q$.
Each has weight at most $\prod_{v\in V(F)}m_v(\delta_F(v))$.
Multiplying~\eqref{eq:holant-support-condition} by
$(Q/s)^{\ell-k}$ therefore gives $W_F\le g_F$.
The two prescribed orientations are distinct, so $B_F^+$ and
$B_F^-$ are disjoint. With $p_F=d_F^++d_F^-$, arithmetic--geometric
mean now gives $W_F\le g_F\le p_F/2$.

\emph{3. Constructing the first-hit witness.}
Start at a nonisolated vertex $a$ and run Tarjan's depth-first search
on the selected directed edges, using fixed incidence orders and
revealing a vertex's outgoing set only when it is first discovered.
Stop immediately upon revealing failure or when the first strongly
connected component is completed and removed from the component
stack. We do not continue to explore other reachable components.
Let $H_F$ be the event of discovering a vertex of $F$ before stopping,
counting discovery before its outgoing set is revealed and a starting
vertex in $F$ as a hit. Set $E_F=H_F\cap(B_F^+\cup B_F^-)$.

At the first hit of $F$, every outgoing set at a vertex of $F$ is
still unexposed. Hence, conditional on any such first-hit history,
the probability of completing either prescribed orientation is $p_F$,
and
\begin{equation}\label{eq:holant-first-hit-completion}
 \Prb_a(E_F)=p_F\Prb_a(H_F),\qquad
 W_F\Prb_a(H_F)\le\tfrac12\Prb_a(E_F).
\end{equation}
On $E_F$, all selected edges from $F$ stay in $F$, and the selected
orientation is strongly connected. The depth-first search therefore
discovers all of $F$ and completes it as one component before
returning to the earlier search. No vertex outside $F$ can join this
component, since $F$ has no outgoing edge to it. Thus the first
completed component is exactly $F$.

The events $E_F$ are consequently mutually exclusive: an exploration
has only one terminal component. Moreover, if $F$ intersects $D$,
discovering all of $F$ also hits $D$, so $E_F\subseteq H_D$.
Group supports with the same vertex set before applying
\cref{pol:lem:first-hit}: for $U\subseteq V(H)$ take total activity
$\sum_{F:V(F)=U}W_F$ and completion event
$\bigsqcup_{F:V(F)=U}E_F$. Apply the lemma with constant $1/2$ and
$\mathcal S$ the nonisolated vertices. It constructs
$h_C=\sum_{a\in\mathcal S}\Prb_a(H_{F_C})$ and gives
$AWh\le h/2$, $1\le h_C\le|\mathcal S|$, and
$\sum_Cw_Ch_C\le|\mathcal S|/2$.
These are the claimed bounds.
\end{proof}

\begin{corollary}[Sampling inputs for the component model]\label{potts:cor:sampling-inputs}
Under the hypotheses of \cref{thm:holant-criterion},
\[
 \lambda_{\min}(AW)\ge-1/2,
 \qquad
 \sum_C w_C e^{D(C)}\le\sqrt e\,|V(H)|.
\]
Thus \cref{pol:lem:sampling} applies with $N=2|V(H)|$, $\eta=1$,
and $\delta=1/2$, and uses $O(|V(H)|\log^2(|V(H)|/\varepsilon))$
steps. The same conclusion holds when each type is split into marks
with the same incompatibility neighborhood.
\end{corollary}
\begin{proof}
The symmetric conflict matrix has nonnegative entries. The first-hit
bound therefore controls its spectral radius by $1/2$, giving the
lower bound as well. For the activity tail, let
$B_F=B_F^+\cup B_F^-$ and $p_F=\Prb(B_F)$ as in the proof of
\cref{thm:holant-criterion}. Events on disjoint vertex supports use
disjoint rows and are jointly independent. Two distinct intersecting
supports cannot both occur: each would be a closed strongly connected
component of the same selected digraph, with all its outgoing edges
specified. At most $|V(H)|$ events occur simultaneously.

Apply \cref{spin:lem:event-bounds} to these events. Since
$W_F=\sum_{C:F_C=F}w_C\le p_F/2$, conflicts depend only on support, and
the conflict neighborhood includes the support itself,
\[
\begin{aligned}
 \sum_C w_Ce^{D(C)}
 &=\sum_F W_F\exp\!\left(\sum_{D:V(D)\cap V(F)\ne\varnothing}W_D\right)\\
 &\le\frac12\sum_Fp_F
       \exp\!\left(\frac12\sum_{D:V(D)\cap V(F)\ne\varnothing}p_D\right)
 \le\sqrt e\,|V(H)|.
\end{aligned}
\]
The first-hit bound gives total activity at most $|V(H)|/2$,
and each compatible family has at most $|V(H)|$ members.
Splitting marks preserves these quantities and the witness inequality.
\end{proof}

\subsection{Region contraction and the temperature range}
\label{sec:flow-coupling}\label{sec:rc-contraction}\label{sec:rc}
\label{sec:regional-implementation}\label{potts:sec:scalar}
We partition $G$ into small connected regions so that the quotient's
sparsity pays for the number of flow labellings in the first-hit
criterion. Summing over each region gives an exact exterior-flow law.
We then bound its signatures and verify the criterion's local inequality.

Set
\begin{equation}\label{eq:regionalparameters}
 B_0=64\lceil\log_2q\rceil,\qquad
 j=\max\{0,\lfloor2B_0/\Delta\rfloor-1\},\qquad \Omega=j/B_0.
\end{equation}

\begin{lemma}[Small regions and a sparse quotient]\label{lem:regions}
There is a partition into connected regions, each with at most
$B_0$ internal edges, whose loopless contraction $H$ has maximum
degree at most $\Delta$ and satisfies
\begin{equation}\label{eq:sparse}
 \Omega e_H(U)\le|U|-1\qquad(\varnothing\ne U\subseteq V(H)).
\end{equation}
Every nonsingleton region has fewer than $\Delta$ exterior edges.
The partition can be constructed in $O(m+nB_0^2)$ work.
\end{lemma}
\begin{proof}
Start with singleton regions and merge a connected set $U$ whenever
$\Omega e_H(U)>|U|-1$. If $j=0$, no merge is needed.
Otherwise put $\eta=2/\Delta-\Omega\ge1/B_0$.
The potential $\Omega e_G(R)-|R|+1$ of every nonsingleton region
is positive: a merge adds the old potentials and the positive
quotient potential. Since $e_G(R)\le \Delta|R|/2$, this gives
\begin{equation}\label{eq:regionsize}
 |R|<\frac{2}{\Delta\eta},\qquad e_G(R)<1/\eta\le B_0,\qquad
 |\delta_G(R)|\le \Delta|R|-2e_G(R)<\Delta.
\end{equation}
Merging therefore preserves the maximum degree bound and terminates
with~\eqref{eq:sparse}. The local cut implementation proving the
work bound is given below.
\end{proof}

\paragraph{The regional signature factors.}
Fixing boundary flow labels prescribes the divergence of the internal
flow. The following coupling expresses its partition function relative
to the zero-divergence case as a probability. This both bounds the
regional signature and implements its acceptance test.

Fix a finite loopless multigraph $H$, an integer $q\ge2$, and
$0<r<1$. For $b\in\mathbb Z_q^{V(H)}$, define the weighted sum
$Z_H(b)=\sum_{\xi\in\Omega_b(H)}r^{|\operatorname{supp}\xi|}$.
The case $b=0$ is the edge-weighted flow model. Nonzero $b$ records
the internal divergence induced when exterior labels are fixed.
Let $\mathcal E_b$ be the event that every open component has total
charge zero: $\sum_{v\in K}b_v=0$ for each component $K$.

\begin{lemma}[Flow--random-cluster coupling with prescribed divergence]\label{lem:flow-completion}
For the random-cluster law $\mu_H$ at activity $p=qr/(1-r)$,
$Z_H(b)/Z_H(0)=\mu_H(\mathcal E_b)$.
If $Z_H(b)>0$, sample $\xi\in\Omega_b(H)$ with weight
$r^{|\operatorname{supp}\xi|}$, include its nonzero edges, and
independently include each zero edge with probability $r$.
The resulting edge law is $\mu_H(\cdot\mid\mathcal E_b)$.
\end{lemma}
\begin{proof}
The number of configurations in $\Omega_b(H)$ vanishing outside $A$
(that is, with $\operatorname{supp}\xi\subseteq A$) is
$q^{|A|-|V(H)|+k(A)}\one_{\mathcal E_b}(A)$: in each component,
choose chord labels freely and solve tree labels; feasibility is
exactly the zero-total-charge condition.
Expanding $r^{\one\{z\ne0\}}=r+(1-r)\one\{z=0\}$ gives edge weight
\begin{equation}\label{eq:completion}
 r^{|A|}(1-r)^{|E(H)|-|A|}
 q^{|A|-|V(H)|+k(A)}\one_{\mathcal E_b}(A)
 \ \propto\ p^{|A|}q^{k(A)}\one_{\mathcal E_b}(A).
\end{equation}
The prefactor is independent of $b$, proving the ratio identity.
Conditional on the flow, the expansion is precisely the stated
independent completion of its zero edges.
\end{proof}

At $b=0$ all flows are conserved and $\mathcal E_b$ is automatic.
This is the flow--random-cluster coupling of~\cite[Section~2.3]{HPR23Flows};
for $q=2$ it is the even-subgraph coupling of~\cite{GrimmettJanson09}.
The lemma also records the version with prescribed divergence.

Apply the identity to the regions constructed above.

For each region $R$, let
$\xi=(\xi_e)_{e\in\delta(R)}$ be its boundary labels viewed outward
from $R$, and put $b_R(\xi)(v)=-\sum_{e\in\delta(R),\ e\ni v}\xi_e$
for each original vertex $v\in R$. Internal labels then have divergence
$b_R(\xi)$, while the complete configuration remains conserved.

\begin{lemma}[RC representation of contracted signatures]\label{lem:regional-signature}
The exterior marginal of the conserved flow law is a Holant on the
loopless contraction, with normalized signatures
\begin{equation}\label{eq:regional-signature}
 f_R(\xi)=r^{|\operatorname{supp}\xi|/2}\chi_R(\xi),\qquad
 \chi_R(\xi)=\frac{Z_{G[R]}(b_R(\xi))}{Z_{G[R]}(0)}\in[0,1].
\end{equation}
These signatures equal one at zero and vanish unless the sum of their
outward boundary labels is zero. Moreover, $\chi_R(\xi)$ is the
probability that every component of a free random-cluster sample on
$G[R]$ has total charge zero.
\end{lemma}
\begin{proof}
Summing the internal flows factors over regions; the two endpoint
half-weights give each exterior edge its factor $r$.
Summing divergence over a region gives boundary conservation.
Lemma~\ref{lem:flow-completion} identifies the remaining ratio as the
stated probability.
\end{proof}

In particular, $\chi_R(\xi)\le1$, so the regional factor is bounded
by its boundary-edge weight. To implement this factor, we enumerate
the $2^{O(\log q)}=q^{O(1)}$ internal edge sets and sample their
random-cluster law. Testing the component charges accepts with
probability exactly $\chi_R(\xi)$. The implementation below retains
these internal edge sets for the final spin recovery.

By Lemma~\ref{lem:regional-signature}, the exact exterior Holant has
$m_R(S)\le r^{|S|/2}$. We use homogeneous incidence parameters.
Write $Q=q-1$ and choose
\begin{equation}\label{eq:rows}
 \kappa=\begin{cases}
 1,&\Delta=3,\\
 \min\{\sqrt Q,(4/3)^\Delta-1\},&\Delta\ge4,
 \end{cases}
 \qquad s=Q/\kappa,\qquad
 t=(s^2/Q)^\Omega,\qquad x=sr/t.
\end{equation}
These choices have $\sqrt Q\le s\le Q$ and $t\ge1$.

\begin{lemma}[The $q\ge3$ certificate]\label{lem:regional-certificate}
For $p\le p_{\od}(q,\Delta)/10$, the contracted Holant satisfies
Theorem~\ref{thm:holant-criterion} with the parameters in
\eqref{eq:rows}. Its outgoing-row mass obeys
\begin{equation}\label{eq:row-slack}
 a:=\frac{(1+\kappa x)^\Delta-1}{\kappa}<1-\frac1{256}.
\end{equation}
\end{lemma}
\begin{proof}
For a connected support with $k\ge2$ vertices and $\ell$ edges,
the forest inequality gives
$s^k/t^\ell\ge Q(Q/s)^{k-2}\ge Q$.
Consequently
\begin{equation}\label{eq:regional-support-budget}
 Qs^{\ell-k}\prod_{R\in V(F)}m_R(\delta_F(R))
 \le Qs^{\ell-k}r^\ell\le x^\ell,
\end{equation}
which is the support condition, including parallel edges.
The local left side is at most $a$.
The scalar estimate below proves
\eqref{eq:row-slack}; this is the only place where the constant
$1/10$ enters.
\end{proof}

\paragraph{Verifying the temperature range.}

We verify the local slack~\eqref{eq:row-slack} used in
Lemma~\ref{lem:regional-certificate}. Keep $Q,\kappa,s$ from~\eqref{eq:rows} and put
\begin{equation}\label{eq:ideal-row-threshold}
 r_0=Q^{2/\Delta-1}\kappa^{-4/\Delta}\big((1+\kappa)^{1/\Delta}-1\big),
 \qquad p_0=\frac{qr_0}{1-r_0}.
\end{equation}
We first prove $r_0<1/3$ and $p_0>(27/256)p_{\od}$, then account
for the rounding of $\Omega$.

\emph{1. The ideal threshold.}
Since $Q\ge\kappa^2$ and $\kappa\ge1$, concavity gives
$r_0\le((1+\kappa)^{1/\Delta}-1)/\kappa^2
\le2^{1/\Delta}-1<1/3$.
For $\Delta=3$, let $C=2^{1/3}-1$ and $z=Q^{-1/3}$.
Direct cancellation gives
$p_0/p_{\od}=C(1-z+z^2)/(1-Cz)>3C/4>3/16>27/256$.

For $\Delta\ge4$, whenever $r_0=CQ^{2/\Delta-1}$, write
\begin{equation}\label{eq:scalarG}
 G:=\frac{qr_0}{p_{\od}}
 =C(1+Q^{-1})\frac{1-Q^{2/\Delta-1}}{1-Q^{-2/\Delta}}\ge C.
\end{equation}
If $\kappa=(4/3)^\Delta-1$, then
$C=\kappa^{-4/\Delta}/3>27/256$, which suffices.
Otherwise $\kappa=\sqrt Q$. Put
$u=(1+\sqrt Q)^{1/\Delta}\le4/3$ and
$b=2\log Q/\log(1+\sqrt Q)<4$.
Now $r_0=(u-1)/Q$ and
\[
 G=\frac{u-1}{u^b-1}(1+Q^{-1})(1-Q^{2/\Delta-1}).
\]
The function $f(Q)=(1+Q^{-1})(1-Q^{-1/2})$ is increasing.
For $Q\ge6$, the ratio $(u^b-1)/(u-1)$ is less than
$1+u+u^2+u^3\le175/27$.
Since $f(6)>175/256$, this gives $G>27/256$.
For $2\le Q\le5$, one has $b<3$, equivalently
$Q^2<(1+\sqrt Q)^3$, so the ratio is at most $37/9$.
Here $f(Q)\ge f(2)=\tfrac32(1-1/\sqrt2)>111/256$,
giving the same bound. Finally $p_0/p_{\od}>G$.

\emph{2. Rounding and uniform slack.}
The exact row-mass threshold at $\Omega$ is
$r_{\rm cap}=(s^2/Q)^\Omega Q^{-1}((1+\kappa)^{1/\Delta}-1)$.
By~\eqref{eq:regionalparameters},
$0<2/\Delta-\Omega<2/B_0$ and $0\le\log(s^2/Q)\le\log q$, so
\[
 \frac{r_{\rm cap}}{r_0}
 =\exp[-(2/\Delta-\Omega)\log(s^2/Q)]>\frac{31}{32}.
\]
Since $p/p_0<128/135$ and $r_0<1/3$, the conversion
$r=p/(q+p)$ gives $r/r_0<192/199$.
The row mass is convex in $r$, zero at zero and one at
$r_{\rm cap}$. Therefore
$a\le r/r_{\rm cap}<6144/6169<1-1/256$, proving
\eqref{eq:row-slack}.

\subsection{Implementation and spin recovery}

We implement the region construction and the exterior-flow proposals,
then recover Potts spins by completing the flow to a random-cluster
sample and coloring its components. The internal marks used in the
proposal tests also supply the regional part of this completion.

\paragraph{Constructing the regions.}
For $j>0$, the following local implementation costs $O(mB_0^2)$.
For one root as source, replace each graph edge by opposite arcs of
capacity $j$, and give each vertex $u$ a sink arc of capacity
$2B_0-j\deg_H(u)\ge \Delta$.
A source-side set $U$ has cut value $2B_0|U|-2j e_H(U)$; a cut below
$2B_0$ is exactly a violating set. Run integral augmenting-path search,
stopping when the flow reaches $2B_0$. At a reached vertex first check
its residual sink arc: if positive, augment immediately; otherwise
expose its spatial neighbours. Sink flows never decrease. At most
$2B_0/\Delta$ vertices can have saturated sink arcs, so at most $2B_0$ spatial
incidences are exposed in total. There are at most $2B_0$ augmentations,
each taking $O(B_0)$ work. If no path remains, the source-reachable set
is the required connected cut. Initially queue every vertex identifier;
after a merge queue only its new identifier. Every violating set always
contains a pending identifier, since sets avoiding a merge are unchanged.
Thus at most $2n-1$ searches suffice. Region lists cost $O(B_0)$ per merge
by~\eqref{eq:regionsize}.

\paragraph{Internal signature tests.}
For each region, enumerate its at most $2^{B_0}$ internal edge sets
and tabulate their free random-cluster law. This costs
$q^{O(1)}m$ work in total. A draw and the charge test of
Lemma~\ref{lem:regional-signature} cost
$\widetilde O(B_0+|\delta_F(R)|)$ for a proposed support $F$.
The test implements $\chi_R(\xi)$ without evaluating a charged
partition function.

\paragraph{A two-orientation proposal.}
Use the outgoing rows in the proof of
Theorem~\ref{thm:holant-criterion}, with the homogeneous parameters
of~\eqref{eq:rows}. Given a root $v$, explore all vertices reachable
from it, sampling each row once. Stop on failure. When exploration
closes, require the entire directed support to be the canonical
depth-first strong orientation of its underlying $F$ or its reversal.
Reject a physical edge selected in both directions and every other
outcome; no transient part is discarded. Recognizing these two
orientations takes $O(|F|)$ work.

For $k=|V(F)|$, $\ell=|F|$, and $b=\ell-k+1$, each orientation
has probability $d_F=\kappa^{\ell-k}x^\ell$.
Assign independent uniform nonzero $\mathbb Z_q$ labels to the
chords of a fixed spanning tree, solve the tree labels by conservation,
and reject zero tree labels. Every valid labelling has probability
$Q^{-b}$. Retain the candidate with probability
\begin{equation}\label{eq:proposal-acceptance}
 \frac{Q^b r^\ell}{2d_F}=\frac{Qt^\ell}{2s^k}\le\frac12,
\end{equation}
then perform the independent signature tests $\chi_R(\xi)$ at its
vertices. Their product gives the remaining factor of $w_C$.
The two orientation events show that each labelled $C$ containing
$v$ is returned with probability exactly $w_C$.
Choosing the root uniformly among the $|V(H)|$ quotient vertices
and retaining the output with probability $1/k$ gives probability
$w_C/|V(H)|$. With $N=2|V(H)|$, running this trial with probability
$1/8$ gives the insertion probability $w_C/(4N)$ required by
\cref{pol:lem:sampling}.

Every fresh row fails with probability at least $1-a>1/256$, so
the number $J$ of queried rows has
$\Prb(J\ge j)\le a^{j-1}$, including failed attempts.
The expected outgoing-list size is $O(1+\log q)$: for
$b_0=\kappa x/(1+\kappa x)$, the row bound gives
$\Delta b_0\le\log(1+\kappa)\le\log q$.
To generate a row at a vertex of degree $d$, first declare success
with probability $a_d=((1+\kappa x)^d-1)/\kappa$ and otherwise fail.
Conditional on success, the outgoing set is a Bernoulli-$b_0$ subset
of the indexed incidence list conditioned to be nonempty.
Its first selected index $K$ has
\[
 \Prb(K\le k\mid K\le d)
 =\frac{1-(1-b_0)^k}{1-(1-b_0)^d},\qquad 1\le k\le d.
\]
For a uniform $U\in[0,1)$, generate this truncated geometric variable by
\[
 K=1+\left\lfloor
 \frac{\log\!\bigl(1-U[1-(1-b_0)^d]\bigr)}{\log(1-b_0)}
 \right\rfloor,
\]
then generate ordinary geometric gaps until the list is exhausted.
There is no rejection of empty subsets and no scan of unselected
incidences. In the model of \cref{sec:computational-model}, the work
is $O(1+|O|)$, with
$\E[|O|\mid\text{success}]=O(1+db_0)=O(1+\log q)$.
The failure probability $1-a$ also bounds the number of queried rows,
so charging recognition, labels, and signature tests to
the exposed rows gives expected proposal work
$\tau=\widetilde O(B_0+1+\log q)$, with an absolute constant.
An owner array tests vertex intersections. The same bound covers the
description maintenance cost $\ell_C=O(|F|+B_0k)$ for maintaining a returned
component and its marks.

\paragraph{Completion and work.}
Keep the internal RC sample $B_R$ whenever a charge test succeeds.
Conditional on the exterior labels, its law is the free regional RC
law conditioned on zero charge in each component. Regard these
marks as part of the polymer type: their weights sum to $w_C$ and
their incompatibility neighborhoods agree, so the witness bound persists.

At output, draw unconditional marks in untouched regions, retain the
stored marks in occupied regions, and include the nonzero exterior
edges. Include each remaining exterior edge independently with
probability $r$. Lemma~\ref{lem:flow-completion} identifies this as
the RC completion, so the map preserves total variation error.
Its work is $\widetilde O(B_0m)$.

\begin{proof}[Proof of \cref{spin:thm:potts}]
Build the regions and their internal random-cluster tables. If the
quotient has no edges, the regional laws already factor and can be
sampled directly. Otherwise, \cref{lem:regional-certificate} verifies
the first-hit criterion, and \cref{potts:cor:sampling-inputs} gives
$O(|V(H)|\log^2(n/\varepsilon))$ insertion--deletion steps.
Each step makes at most one of the preceding trials. Its expected
cost, including a returned component's description and signature
marks, is $\widetilde O(B_0+\log q)$. An owner array tests conflicts
and records insertions; deletion work is charged to the insertion of
that component. Thus the chain costs
$\widetilde O(|V(H)|(B_0+\log q))$ expected work.

The regional preprocessing costs $q^{O(1)}m$ and the final
random-cluster completion costs $\widetilde O(B_0m)$. Coloring its
components takes $O(m)$ further work and produces Potts spins.
All projection and recovery kernels contract total variation.
Since $B_0=O(\log q)$ and $|V(H)|\le n\le2m$ on the reduced graph,
its total work is $\widetilde O(q^{O(1)}m)$.
Adding preprocessing and the independent isolated spins gives both
stated guarantees, with an absolute exponent of $q$.
\end{proof}

\section{Discussion and open problems}\label{sec:discussion}

Polynomial-time tractability of the ferromagnetic Ising model has
been known since \cite{JSIsing93,RW99}. Our result strengthens the
sampling guarantee to near-linear work throughout the tree-uniqueness
regime, including the threshold where single-site Glauber dynamics
can require $\Omega(n^{3/2})$ updates \cite{CCYZ25}.
Even Swendsen--Wang dynamics requires $\Omega(n^{1/8})$ full sweeps
on some critical cubic graphs (\cref{sw:prop:cubic}), giving a
superlinear work bound for ordinary sweeps run to worst-case mixing.
The cycle representation retains a fixed lower spectral margin at
that boundary, and its exploration makes the resulting updates
efficient to implement. The XY and Potts applications show how the
same polymer criterion can be used with more general flow types.
The first two questions concern deterministic counting at the Ising
threshold and the full Potts coexistence range. We then discuss
extensions and sharper sampling guarantees.

\paragraph{A deterministic FPTAS for critical Ising.}
For every fixed $\Delta\ge3$, is there a deterministic FPTAS for
the zero-field ferromagnetic Ising partition function at
$\beta=\Delta/(\Delta-2)$ on all graphs of maximum degree at most
$\Delta$? Concretely, given such a graph and $0<\varepsilon<1$,
the algorithm should return $\widehat Z$ satisfying
\[
 (1-\varepsilon)Z_{\rm Ising}
 \le\widehat Z\le
 (1+\varepsilon)Z_{\rm Ising}
\]
in time polynomial in $n$ and $1/\varepsilon$.
The zero-free approach of \cite{LSS19Fisher} gives an FPTAS
strictly inside the uniqueness regime, with constants depending
on the distance to its boundary. Extending deterministic counting
to the endpoint therefore requires more than substituting the
critical parameter into that bound.
Our cycle representation has a fixed lower spectral margin even
there, but this controls a randomized chain rather than a
deterministic approximation to its partition function.
Can the completion-event structure yield a deterministic truncation
or recursion with polynomial cost at criticality?
A stronger goal is a polynomial exponent independent of $\Delta$.

\paragraph{Near-linear sampling for other critical models.}
For each fixed $\Delta\ge3$, can the hard-core model at
$\lambda_c(\Delta)=(\Delta-1)^{\Delta-1}/(\Delta-2)^\Delta$,
or the zero-field antiferromagnetic Ising model at edge activity
$\beta=(\Delta-2)/\Delta$, be sampled on every graph of maximum
degree at most $\Delta$ in $\widetilde O(n)$ work?
Even an $n^{1+o(1)}$ bound at inverse-polynomial total-variation
error would improve the known polynomial guarantees.
The worst-case Glauber mixing lower bounds
$\Omega(n^{4/3})$ and $\Omega(n^{3/2})$, respectively
\cite{CCYZ25}, concern that chain rather than arbitrary samplers.
For zero-field antiferromagnetic Ising on a bipartite graph,
flipping all spins on one side changes the edge activity to
$1/\beta$, so \cref{spin:thm:ising} already gives near-linear
sampling at criticality. The question is therefore to handle
general graphs as well. On such graphs, odd cycles give signed
weights in the antiferromagnetic high-temperature expansion,
preventing a direct use of our nonnegative cycle representation.

\paragraph{Near-linear Ising sampling beyond the tree threshold.}
Although Feng, Guo, and Zhang \cite[Section~6]{FGZ26ZeroField}
conjecture that random-cluster Glauber dynamics does not mix
in near-linear time in general, our algorithm achieves
near-linear sampling at the tree-uniqueness threshold by
bypassing the random-cluster Glauber dynamics.
The exploration underlying our method, however, relies on
$(\Delta-1)(\beta-1)/(\beta+1)\le1$, the nonbacktracking transition weights
summing to at most one, so the present proposal construction
and spectral certificate no longer apply.
This motivates the search for a new sampling method:
can the zero-field ferromagnetic Ising model on every simple graph
be sampled in $\widetilde O_\beta(n+m)$ expected work at
inverse-polynomial accuracy for every fixed $\beta>1$?

\paragraph{Disclosure of AI use.}
The main ideas of this work were developed with assistance from OpenAI's
Codex. The human author contributed original ideas in developing the
AI-generated initial drafts and through continued interaction with the
AI. The author worked through the proofs in detail and refined, restructured,
and rewrote the arguments.
AI assistance was also used in preparing, auditing, and revising the
manuscript. The author takes full responsibility for its mathematical
content and final presentation.

\begingroup
\small
\setlength{\emergencystretch}{2em}
\raggedright
\urlstyle{same}
\AddToHookNext{env/thebibliography/begin}{%
  \phantomsection
  \addcontentsline{toc}{section}{References}%
  \setlength{\itemsep}{2pt}%
}
\bibliographystyle{plainurl}
\bibliography{references,matching2-spins}
\endgroup

\appendix

\section{General-Temperature XY Sampling via Permanents}
\label{spin:app:permanent}

The algorithm in \cref{spin:sec:xy} is designed for almost-linear work
in the tree-threshold regime. This appendix gives polynomial-time
counting and sampling at arbitrary ferromagnetic couplings, using a
finite nonnegative matrix. The reduction includes a quantitative
finite-capacity error bound, so no limiting object is needed by the
algorithm.

Allow edge-dependent $\beta_e\ge1$ and write $K_e=\log\beta_e$.
With normalized Haar measure, the partition function is
\[
 Z_{\rm XY}(G,\mathbf K)=
 \int\exp\!\left(\sum_{uv\in E}K_{uv}\cos(\theta_u-\theta_v)\right)
       \prod_{v\in V}\frac{d\theta_v}{2\pi}.
\]

\begin{theorem}[XY tractability at every fixed temperature]\label{spin:thm:xy-general}
For $K_e\ge0$ and $0<\varepsilon<1/2$, there are randomized algorithms
that approximate $Z_{\rm XY}(G,\mathbf K)$ to relative error
$\varepsilon$ with probability at least $3/4$, and sample its spin law
within total variation $\varepsilon$. Their work is polynomial in
$n,m,\sum_eK_e$, and $1/\varepsilon$ in the computational model of
\cref{sec:computational-model}. In particular, both tasks take
polynomial work at every fixed uniform $\beta\ge1$.
\end{theorem}

We use nonnegative permanent approximation as a black box. The
polynomial-time result of Jerrum, Sinclair, and Vigoda \cite{JSV04}
has been sharpened to a $\widetilde O(d^3\varepsilon^{-2})$ arithmetic
bound for a $d\times d$ matrix by Chen, Guo, Vigoda, and Yang
\cite[Corollary~2]{CGVY26Permanent}. Polynomial-time approximate
weighted perfect-matching sampling follows as well, either from their
sampling construction or from counting by self-reduction
\cite{JVV86}. We need only this counting and sampling interface.

For each directed edge $a$ put $t_a=K_{e(a)}/2$.
A directed current is a family $N=(N_a)_{a\in\vec E}$ of nonnegative
integers. It is balanced when its total incoming and outgoing values
agree at every vertex; write this common value as
$d_v=\sum_{a:\operatorname{tail}(a)=v}N_a$.
Expanding
$e^{K\cos\theta}=e^{(K/2)e^{i\theta}}e^{(K/2)e^{-i\theta}}$
and integrating the vertex angles gives
\begin{equation}\label{perm:eq:directed-current}
 Z_{\rm XY}=\sum_{\operatorname{div}N=0}
                 \prod_{a\in\vec E}\frac{t_a^{N_a}}{N_a!}.
\end{equation}
Let $\nu$ be the normalized directed-current law in this identity.

For an integer $L\ge1$, define a matrix $B_L$ with rows and columns
indexed by $(v,i)\in V\times[L]$:
\begin{equation}\label{perm:eq:matrix}
 B_L[(u,i),(v,j)]=
 \begin{cases}
 1,&u=v,\\
 K_{uv}/(2L),&uv\in E,\\
 0,&\text{otherwise}.
 \end{cases}
\end{equation}
Thus each diagonal vertex block is an all-ones matrix. Put
$(L)_d=L!/(L-d)!$ for $d\le L$ and $(L)_d=0$ for $d>L$, and define
$a_L(N)=\prod_v (L)_{d_v}/L^{d_v}$.

\begin{proposition}[Finite-capacity identity]\label{perm:prop:identity}
For every $L\ge1$,
\begin{equation}\label{perm:eq:identity}
 \frac{\operatorname{perm}(B_L)}{(L!)^n}
 =\sum_{\operatorname{div}N=0}
     \left(\prod_a\frac{t_a^{N_a}}{N_a!}\right)a_L(N)
 =Z_{\rm XY}\E_\nu a_L.
\end{equation}
The block counts of a weighted perfect matching of $B_L$ have law
$\nu_L(N)=a_L(N)\nu(N)/\E_\nu a_L$.
\end{proposition}
\begin{proof}
Fix the off-diagonal block counts $N_{u\to v}$ of a perfect matching.
The number of unused left and right copies at each vertex must be
equal, so these counts are balanced. For a balanced $N$ with
$d_v\le L$, choose the outgoing and incoming copies in each block
and match the selected copies. The resulting number of perfect
matchings is
\[
 \frac{(L!)^{2n}}
 {\prod_v(L-d_v)!\prod_aN_a!}.
\]
Indeed, the ordered partitions of left and right copies contribute
$(L!)^{2n}/[\prod_v((L-d_v)!)^2\prod_a(N_a!)^2]$;
the cross-block bijections contribute $\prod_aN_a!$ and the
within-block bijections contribute $\prod_v(L-d_v)!$.
Multiplying by $\prod_a(t_a/L)^{N_a}$ and dividing by $(L!)^n$
gives the summand in \eqref{perm:eq:identity}, since
$\sum_aN_a=\sum_vd_v$. Summing proves the partition-function and
distribution identities.
\end{proof}

\begin{lemma}\label{perm:lem:error}
For every $L\ge1$,
\[
 0\le1-\E_\nu a_L\le\frac{(\sum_eK_e)^2}{2L},
 \qquad
 \TV(\nu_L,\nu)\le1-\E_\nu a_L.
\]
\end{lemma}
\begin{proof}
The falling-factorial product gives the pointwise bound
\begin{equation}\label{perm:eq:capacity-loss}
 1-a_L(N)\le\frac1{2L}\sum_vd_v(d_v-1)
           \le\frac{T(T-1)}{2L},\qquad T=\sum_aN_a.
\end{equation}
For $d_v\le L$ this follows by expanding the product of factors
$1-j/L$ and using the union bound. If any $d_v>L$, the right side
is at least one.

For an integer divergence vector $b$, let $Z_b$ be the current sum
with $\operatorname{div}N=b$. Fourier expansion expresses $Z_b/Z_0$
as a Fourier moment of the XY spin law. Its modulus is at most one,
and its current coefficients are nonnegative, so $0\le Z_b/Z_0\le1$.
Let $\partial a$ be the divergence of one unit on directed edge $a$.
Removing two specified units gives the factorial-moment identity
\[
 \E_\nu[N_a(N_b-\one_{\{a=b\}})]
 =t_at_b\frac{Z_{-\partial a-\partial b}}{Z_0}
 \le t_at_b.
\]
Summing over directed edges gives
$\E_\nu[T(T-1)]\le(\sum_at_a)^2=(\sum_eK_e)^2$.
Together with \eqref{perm:eq:capacity-loss}, this proves the first bound.

For the second, put $A=\E_\nu a_L>0$. Since $0\le a_L\le1$,
$(A-a_L)_+\le A(1-a_L)$. Hence
\[
 \TV(\nu_L,\nu)=\frac{\E_\nu(A-a_L)_+}{A}
 \le1-A.
\]
\end{proof}

\begin{proof}[Proof of \cref{spin:thm:xy-general}]
Take $L=\max\{1,\lceil4(\sum_eK_e)^2/\varepsilon\rceil\}$.
By \cref{perm:prop:identity,perm:lem:error}, approximating
$\operatorname{perm}(B_L)$ to relative error $\varepsilon/4$ and
dividing by $(L!)^n$ approximates $Z_{\rm XY}$ to relative error
$\varepsilon$. The permanent algorithm provides success probability
at least $3/4$.

For sampling, obtain a weighted perfect matching of $B_L$ within
TV distance $\varepsilon/4$ and project to its off-diagonal block
counts. The output law is within $\varepsilon/4$ of $\nu_L$, and
$\TV(\nu_L,\nu)\le\varepsilon/8$.

Apply the directed-current stage of the recovery kernel in
\cref{spin:lem:xy-recovery}: pair incoming and outgoing links
uniformly, forget directions, color loops with two colors, and
sample the beta amplitudes $u_v$. With nonuniform couplings the
conditional Ising activities are
\[
 \beta^1_{uv}=\exp(2K_{uv}\sqrt{u_uu_v}),\qquad
 \beta^2_{uv}=\exp(2K_{uv}\sqrt{(1-u_u)(1-u_v)}).
\]
The same beta integral in that proof gives the joint law, edge by
edge. The current-plus-Bernoulli completion and cluster signs then
return XY spins. No Bessel lifting is needed because the input is
already a directed current. This kernel contracts TV distance and
uses $O(m+T)$ work on a current with $T$ units.

The matrix has dimension $nL$, and the permanent counting and sampling
procedures use polynomial work in that dimension and $1/\varepsilon$.
For every projected matching, $T=\sum_vd_v\le nL$, so recovery also
takes polynomial work. This proves the stated dependence on
$n,m,\sum_eK_e$, and $1/\varepsilon$.
\end{proof}

The exact identity also implies
$Z_{\rm XY}=\lim_{L\to\infty}\operatorname{perm}(B_L)/(L!)^n$,
since $a_L(N)\to1$ and $0\le a_L\le1$. The algorithms above use
the finite value of $L$ fixed by the error bound.

\section{A Swendsen--Wang Mixing Lower Bound at the Tree Threshold}
\label{sw:app:lower-bound}

\begin{definition}[Swendsen--Wang dynamics
{\cite[Section~2.2]{GJ18RandomCluster}}]\label{sw:def:dynamics}
Let $G=(V,E)$ and $\beta>1$. Set $p=1-\beta^{-1}$ and
$E(\sigma)=\{uv\in E:\sigma_u=\sigma_v\}$ for
$\sigma\in\{-1,1\}^V$.
The Swendsen--Wang (SW) dynamics is the Markov chain on
$\{-1,1\}^V$ with transition kernel
\[
 P_{\rm SW}(\sigma,\tau)
 =\sum_{A\subseteq E(\sigma)\cap E(\tau)}
 p^{|A|}(1-p)^{|E(\sigma)|-|A|}\,2^{-k(A)},
\]
where $k(A)$ is the number of connected components of $(V,A)$,
including isolated vertices. One transition is called a round.
\end{definition}

Equivalently, retain each edge of $E(\sigma)$ independently with
probability $p$, then give all vertices in each retained component
a common uniform spin in $\{-1,1\}$, independently across components.
The chain is reversible with respect to the zero-field Ising law
$\mu_{\rm Ising}(\sigma)\propto\beta^{|E(\sigma)|}$.

We give a polynomial lower bound on the number of rounds on sparse
graphs at the exact tree threshold. The exponent is not optimized.

\begin{proposition}\label{sw:prop:cubic}
For every sufficiently large even $n$, there is a simple three-regular
graph $G_n$ on $n$ vertices such that, for the zero-field Ising law
$\mu_G(\sigma)\propto\beta^{m_{=}(\sigma)}$, $\beta=\beta_c(3)=3$, 
\[
 \gap(P_{\rm SW})=O(n^{-1/8}),\qquad
 \tmix(P_{\rm SW},1/4)=\Omega(n^{1/8}).
\]
A uniform simple cubic graph satisfies these bounds with probability
bounded below by an absolute positive constant.
\end{proposition}

Here $\beta=3=\Delta/(\Delta-2)$ for $\Delta=3$.
An ordinary full sweep uses $\Theta(n)$ work on this graph family,
so the proposition gives an $\Omega(n^{9/8})$ work obstruction for
such sweeps run to worst-case mixing. It is a statement about this
chain and cost model, not a lower bound for every sampler or every
choice of initial distribution.

\begin{proof}
Write $M(\sigma)=\sum_v\sigma_v$ and set
\[
 g(x)=\max\{0,\min\{x-1,3-x\}\},\qquad
 f(\sigma)=g(|M(\sigma)|/n^{3/4}).
\]
Thus $0\le f\le1$, $g$ is $1$-Lipschitz, and $f$ vanishes outside
$I_n=\{\sigma:n^{3/4}<|M(\sigma)|<3n^{3/4}\}$.
This compact support avoids requiring quantitative magnetization
estimates outside the critical window.

\paragraph{Reduction to a planted one-step estimate.}
Let $\Prb_0$ be the cubic configuration model, obtained by uniformly
pairing $3n$ labeled half-edges, and put
\[
 Z_G=\sum_\sigma3^{m_{=}(\sigma)},\qquad
 W_G=\sum_{\sigma\in I_n}3^{m_{=}(\sigma)}.
\]
For each fixed $\sigma$, give the graph the planted density
$3^{m_{=}(\sigma)}/\E_0 3^{m_{=}(\sigma)}$ relative to $\Prb_0$.
The graph is then kept fixed during the SW update.
We will prove, uniformly over $\sigma\in I_n$, that
\begin{equation}\label{sw:eq:step}
 \E_\sigma^*[(f(\sigma')-f(\sigma))^2]=O(n^{-1/8}),
\end{equation}
where the expectation includes the planted graph and one SW round.
Reversibility on each graph and the support of $f$ imply
\begin{equation}\label{sw:eq:energy}
 \E_0[Z_G\cE_G(f,f)]\le Cn^{-1/8}\E_0W_G.
\end{equation}
Indeed, any nonzero term in the Dirichlet sum has an endpoint in
$I_n$. Symmetry allows restriction of the first endpoint to $I_n$
after removing the factor $1/2$; planting then gives
\eqref{sw:eq:energy}.

We use three static estimates of Prodromidis and Sly
\cite{PS26Magnetization}. Their Theorem~1.1 gives the limiting
density proportional to $\exp(-x^4/54)$ for $M/n^{3/4}$ on simple
cubic graphs. Hence $\Var_{\mu_G}f\ge v>0$ with probability $1-o(1)$
conditional on simplicity. Their Lemma~3.2 bounds the second-moment
ratio of each magnetization-restricted partition function uniformly
for $|M|\le3n^{3/4}$. Minkowski's inequality therefore gives
$\E_0W_G^2\le C(\E_0W_G)^2$.
Their Proposition~3.3 gives independent Poisson limits of means
$3/2$ and $5/4$ for planted loops and double edges, respectively.
Consequently,
\[
 \E_0[W_G\one_{\{G\ {\rm simple}\}}]
 =(e^{-11/4}+o(1))\E_0W_G.
\]
These statements use exponential coupling $J=\frac12\log3$ in
their convention. Uniformity follows because their results apply
to every sequence of magnetizations in this window; spin-flip
symmetry covers negative magnetizations. For the simplicity event,
planting a fixed configuration or planting its magnetization gives
the same probability by vertex relabeling.

Let $\mathcal G=\{G\text{ simple},\ \Var_{\mu_G}f\ge v\}$.
Cauchy--Schwarz and the second-moment bound give
$\E_0[W_G\one_{\mathcal G}]\ge c\E_0W_G$.
By \eqref{sw:eq:energy} and $W_G\le Z_G$,
\[
 \E_0[W_G\one_{\{\cE_G(f,f)>Kn^{-1/8}\}}]
 \le (C/K)\E_0W_G.
\]
For a sufficiently large constant $K$, the event
$\mathcal H=\mathcal G\cap\{\cE_G(f,f)\le Kn^{-1/8}\}$
has $W_G$-weighted mass at least $c'\E_0W_G$.
Another application of Cauchy--Schwarz gives
$\Prb_0(\mathcal H)\ge(c'\E_0W_G)^2/\E_0W_G^2>c''$.
Conditioning on simplicity yields the claimed positive probability.
The Rayleigh quotient on $\mathcal H$ bounds the gap by
$(K/v)n^{-1/8}$, and \eqref{mc:eq:gap-lower} gives the
mixing assertion.

\paragraph{The retained graph under planting.}
It remains to prove \eqref{sw:eq:step}.
By symmetry take $M=ns>0$, where $s\asymp n^{-1/4}$, and let
$N_\pm=n(1\pm s)/2$. If $k$ counts cross-spin edges and
$A=3N_+$, $B=3N_-$, its planted weight and successive ratio are
\[
 u(k)=\binom Ak\binom Bk k!(A-k-1)!!(B-k-1)!!3^{-k},
 \qquad
 \frac{u(k+2)}{u(k)}
 =\frac{(A-k)(B-k)}{9(k+1)(k+2)}.
\]
The mode is within $O(1)$ of
$k_*=(3n/16)(\sqrt{9-8s^2}-1)$.
Near the mode, the logarithm of this ratio decreases at rate
$\Theta(1/n)$; multiplying ratios and using their monotonicity
outside that neighborhood gives
\[
 \Prb_\sigma^*(|k-k_*|>\sqrt n\log n)
 \le C\exp(-c\log^2 n).
\]
SW retains same-spin edges with probability $2/3$.
Their numbers satisfy
$L_\pm\mid k\sim\operatorname{Bin}((3N_\pm-k)/2,2/3)$.
Conditional on $L_\pm$, the retained half-edges in each color class
form a uniform $2L_\pm$-subset, paired uniformly.
Binomial concentration and bounded differences for sampling without
replacement therefore give, with failure $C\exp(-c\log^2n)$,
\begin{equation}\label{sw:eq:counts}
 \begin{split}
 n_{\pm,j}
 &=N_\pm\binom3j\alpha_\pm^j(1-\alpha_\pm)^{3-j}
   +O(\sqrt n\log n),\qquad 0\le j\le3,\\
 \alpha_\pm
 &=\frac23\left(1-\frac{k_*}{3N_\pm}\right)
   =\frac12\pm\frac{s}{6}+O(s^2).
 \end{split}
\end{equation}
For example, swapping one selected half-edge changes a degree count
by at most two, and the hypergeometric expectation differs from the
binomial expression by $O(1)$.
Conditional on the retained degrees, the two color graphs are
independent uniform configuration models.

\paragraph{Quantitative exploration of the majority graph.}
Condition on degree counts satisfying \eqref{sw:eq:counts}, and put
$\eta=n^{-1/4}$ and $h=n^{-1/16}$.
Use the exponential-clock exploration of
\cite[Section~5.1]{HJL19Components}, setting aside isolated vertices.
Let $L(t)$ count living half-edges and $S(t)$ actual sleeping
half-edges. The artificial sleeping degree-$j$ count
$\widetilde V_j(t)$ has mean $n_je^{-jt}$, where $n_j=n_{+,j}$.
Write
\[
 \widetilde A(t)=L(t)-\sum_jj\widetilde V_j(t),\qquad
 D(t)=\sum_jj\widetilde V_j(t)-S(t).
\]
The actual active count is $\widetilde A+D$. With
$\ell=\sum_jjn_j$ and $x=e^{-t}$, its artificial deterministic curve is
\[
 a(t)=\ell e^{-2t}-\sum_jjn_je^{-jt}
     =x(1-x)(3n_3x-n_1).
\]
Its positive zero is $t_0=\log(3n_3/n_1)\asymp\eta$.
The degree laws converge to $\operatorname{Bin}(3,1/2)$, so the
assumptions of \cite[Lemma~5.13]{HJL19Components} hold.
That lemma, together with the individual death-process martingale
bounds and $\E L(t)=\ell e^{-2t}+O(1)$, gives, for fixed large $C_0$,
\[
 \E\sup_{t\le C_0\eta}|\widetilde A(t)-a(t)|^2
 +\sum_{j=1}^3\E\sup_{t\le C_0\eta}
   |\widetilde V_j(t)-n_je^{-jt}|^2
 \le C(n\eta+1).
\]
Thus all these errors are at most $r=c_0hn\eta^2$ except with
probability
$O((h^2n\eta^3)^{-1})=O(n^{-1/8})$.
The restart inequality \cite[(5.4)]{HJL19Components} gives
\[
 0\le D(t)\le-\inf_{u\le t}\widetilde A(u)+3;
\]
moreover $D$ does not increase while a component is explored.

On this concentration event, the factorization of $a$ gives
$\widetilde A>0$ throughout
$[C_1h\eta,t_0-C_1h\eta]$, for a suitable constant $C_1$.
A single component $C_*$ spans this interval.
Before its start, $a\ge0$, so $D\le r+3$.
For sufficiently large $C_2$,
$a(t_0+C_2h\eta)<-3r-6$; hence that component must end by
$t_0+C_2h\eta$.
Let $T$ be the first component completion after $t_0-C_1h\eta$,
or the terminal exploration time if there is none.
This is a stopping time. The preceding estimates imply
\[
 |C_*|=N_+-\sum_{j=0}^3n_je^{-jt_0}+O(hn^{3/4}),\qquad
 \#\{\text{vertices explored before }C_*\}=O(hn^{3/4}).
\]
Here actual and artificial sleeping counts differ by at most $D$
for every positive degree, and agree for degree zero.
For exact binomial degrees the main term is $N_+\chi(\alpha_+)$,
where $\chi(\alpha)=1-((1-\alpha)/\alpha)^3$.
The errors in \eqref{sw:eq:counts} change $t_0$ by
$O(n^{-1/2}\log n)$ and this main term by $O(\sqrt n\log n)$.
Expanding $\alpha_+$ therefore gives
\begin{equation}\label{sw:eq:size}
 |C_*|=N_+\chi(\alpha_+)+O(hn^{3/4})
      =M+O(hn^{3/4}).
\end{equation}
The unexplored degree counts at $T$ are
$n_je^{-jt_0}+O(hn\eta)$.
Define excess degree as $\sum_jj(j-1)n_j/\sum_jjn_j-1$.
For the ideal residual binomial counts it equals $1-2\alpha_+$;
the perturbation is $O(h\eta+n^{-1/2}\log n)$.
Thus the residual graph has $\Theta(n)$ half-edges and excess at
most $-c\eta$.
These size and degree conclusions define a good event measurable
at $T$, with failure $O(n^{-1/8})$. Conditional on this history,
the remaining pairing is uniform. We do not condition the
remaining graph on the future maximal-deviation event.

\paragraph{Residual components and recoloring.}
For a configuration model with degrees at most three,
$\Theta(n)$ half-edges, and excess at most $-c\eta$, counting simple
paths of length $j$ from a vertex gives
\[
 \E[\#\{\text{such paths}\}]
 \le C(1-c\eta)^{j-1}\exp(Cj^2/n)
 \le Ce^{-c'\eta j},
 \qquad 1\le j\le c''\eta n.
\]
Indeed, the pairing denominators are
$(\ell-1)(\ell-3)\cdots(\ell-2j+1)$; dropping distinctness of
internal vertices bounds their choices by
$(\sum_i d_i(d_i-1))^{j-1}$.
Summing the path bounds counts vertices at distance below
$j_0=\lfloor c''\eta n\rfloor$.
More distant vertices contribute at most
$n\Prb(\text{a path of length }j_0)\le Cne^{-c'''n\eta^2}$.
Since $\eta=n^{-1/4}$, the expected component size is $O(1/\eta)$,
and the expected sum of squared component sizes is $O(n/\eta)$.
Apply this estimate conditionally to the residual majority graph
at $T$, and directly to the minority graph.
The components explored before $C_*$ have squared-size sum at most
$O(h^2n^{3/2})$ by their total size.

Let $R$ be the sum of squared sizes of all retained components
except $C_*$. Including the degree-count event in the
history-measurable good event $\mathcal B$, we have
\[
 \Prb_\sigma^*(\mathcal B^c)=O(n^{-1/8}),\qquad
 \E_\sigma^*[R\one_{\mathcal B}]
 =O(h^2n^{3/2}+n^{5/4}).
\]
Independent component recoloring gives
$\E[(|M(\sigma')|-|C_*|)^2\mid\text{components}]\le R$.
Together with \eqref{sw:eq:size}, boundedness of $g$, and its
Lipschitz property, this yields
\[
 \E_\sigma^*[(f(\sigma')-f(\sigma))^2]
 \le C(n^{-1/8}+h^2+n^{-1/4})
 =O(n^{-1/8}).
\]
This proves \eqref{sw:eq:step} and completes the proof.
\end{proof}

\end{document}